\documentclass{aa}  
\usepackage{graphicx}
\usepackage{amssymb}

\usepackage{natbib}

\usepackage{booktabs}
\usepackage{multirow}
\usepackage{pdflscape}
\usepackage{subfig}
\usepackage{color}
\usepackage{xcolor}

\usepackage{float}
\usepackage{placeins}
\usepackage{caption}
\usepackage{wrapfig}

\graphicspath{{figures/}}

\offprints{cguepin@umd.edu}

\title{Signatures of secondary acceleration in neutrino flares}
\titlerunning{Signatures of secondary acceleration in neutrino flares}
\date{\today}
\author{Claire Gu\'epin\inst{1,2}}
\authorrunning{C. Gu\'epin}

\institute{Department of Astronomy, University of Maryland, College Park, MD 20742, USA
\and Joint Space-Science Institute, University of Maryland, College Park, MD 20742, USA} 
		     
\abstract{High-energy neutrino flares are interesting prospective counterparts to photon flares, as their detection would guarantee the presence of accelerated hadrons within a source, provide precious information about cosmic-ray acceleration and interactions, and thus impact the subsequent modeling of non-thermal emissions in explosive transients. In these sources, photomeson production can be efficient, producing a large amount of secondary particles, such as charged pions and muons, that decay and produce high-energy neutrinos. Before their decay, secondary particles can experience energy losses and acceleration, which can impact high-energy neutrino spectra and thus affect their detectability. In this work, we focus on the impact of secondary acceleration. We consider a one zone model, characterized mainly by a variability timescale $t_{\rm var}$, a luminosity $L_{\rm bol}$, a bulk Lorentz factor $\Gamma$. The mean magnetic field $B$ is deduced from these parameters. The photon field is modeled by a broken power-law. This generic model allows to evaluate systematically the maximum energy of high-energy neutrinos in the parameter space of explosive transients, and shows that it could be strongly affected by secondary acceleration for a large number of source categories. In order to determine the impact of secondary acceleration on the high-energy neutrino spectrum and in particular on its peak energy and flux, we complement these estimates by several case studies. We show that secondary acceleration can increase the maximum neutrino flux, and produce a secondary peak at the maximum energy in the case of efficient acceleration. Secondary acceleration could therefore enhance the detectability of very-high-energy neutrinos, that will be the target of next generation neutrino detectors such as KM3NeT, IceCube-Gen2, POEMMA or GRAND.}


\begin{document}

\renewcommand\d{\mathrm{d}}

\maketitle
\section{Introduction}

The observations of explosive transients, strengthened by the high sensitivity and time resolution of current instruments, shed light on a great diversity of astrophysical phenomena. The variety of information carried by photons, cosmic rays, neutrinos and gravitational waves help to better understand the physical properties of astrophysical sources, including energetic transients. The recent detection of gravitational waves \citep{LIGO2016a, LIGO2016b}, in particular associated with the merger of two neutron stars \citep{LigoVirgoNSNS17}, has illustrated the power of the multimessenger approach. In this picture, high-energy neutrinos ($>10^{12}\,{\rm eV}$) play an important role, as their detection provides the evidence of hadron acceleration and interactions. The association of high-energy neutrinos with astrophysical sources would therefore help to identify various classes of cosmic-ray accelerators, and notably the sources of cosmic rays above $10^{15}\,{\rm eV}$, that still need to be clearly identified among various theoretical candidates, as for instance gamma-ray bursts, active galactic nuclei, tidal disruption events or new-born pulsars. Since the first detection of high-energy astrophysical neutrinos \citep{Aartsen13a}, the IceCube collaboration has detected more than a hundred of cosmic neutrinos between $10^{13}\,{\rm eV}$ and $10^{16}\,{\rm eV}$. Moreover, methods for time-variable searches of neutrinos have been developed \citep[e.g.][]{Abbasi12a, Aartsen13b, Aartsen15}. The first hints of high-energy neutrino coincidence with blazar flares \citep{IceCube18} highlight the importance of observational and theoretical coincidence studies. By increasing the sensitivity and the accessible energy range, the next generation of high-energy and very-high energy neutrino detectors such as IceCube-Gen2 \citep{IceCubeGen2_2015}, or for instance the Probe of Extreme Multi-Messenger Astrophysics \citep[POEMMA:][]{POEMMA17}, Giant Radio Array for Neutrino Detection \citep{GRAND18}, Trinity \citep{Trinity19} or Radio Neutrino Observatory \citep[RNO:][]{RNO19} concepts, will be decisive for the association of high-energy neutrinos with astrophysical sources.

The peak flux and peak energy of high-energy neutrino flares determine the optimal instruments for their detection. Their maximum energy is related to the energy of accelerated protons, taking into account the energy losses of secondaries, namely charged pions, muons and kaons \citep[e.g.][]{Waxman97_GRB}. Acceleration of secondaries is less frequently mentioned, although it could modify the neutrino spectrum, produce a peak at higher energies, and thus impact the detectability of high-energy neutrino sources. Among the studies considering the acceleration of secondaries, \citet[][Appendix A]{Murase12} examined stochastic acceleration of secondaries by turbulence in gamma-ray bursts. \cite{Klein13} considered linear acceleration applicable to various sources classes. \cite{Reynoso14} and \cite{Winter14} studied a two-zone model with gamma-ray bursts properties, with an acceleration zone and a radiation zone. In this latter case, secondaries diffuse in the radiation zone and can move back into the acceleration zone, where they are accelerated. Given the variety of potential sources of high-energy neutrinos, and the aforementioned importance of coincidence studies, it seems timely to study secondary acceleration in a general model, applicable to numerous categories of explosive transients.

In this work, we assess systematically the impact of secondaries acceleration on the detectability of high-energy neutrino flares in coincidence with photon flares, for various categories of explosive transients. We consider a one zone model, where acceleration and radiation processes can take place. In keeping with the general approach presented in \cite{Guepin17}, we describe explosive transients with a handful of parameters: the distance from the source $D_{\rm s}$, the total luminosity of the source measured during the flare $L_{\rm bol}$, the variability timescale of the emission $t_{\rm var}$, and the bulk Lorentz factor of the outflow $\Gamma$ or the Doppler factor $\delta = 1/[\Gamma(1-\beta \cos \theta)]$. In the following, we consider an emission towards the observer, such as $\theta=0$. The radiation background associated with the observed photon flare serves as a target for photohadronic interactions producing neutrinos. In Section~\ref{sec:model}, we describe the one-zone model and the energy-gain and energy-loss processes considered. In Section~\ref{sec:Emax}, we compare the maximum energy of high-energy neutrinos obtained without or with acceleration of secondaries. In Section~\ref{sec:Spec}, we give benchmark examples of high-energy neutrino spectra for several source categories, in order to highlight the impact of secondary acceleration on high-energy neutrino detectability. We summarize and discuss the implication of our results in Section~\ref{sec:discussion}.

\section{One-zone model - general aspects}
\label{sec:model}

In the following, all primed quantities are in the comoving frame of the flaring region, and other quantities are in the observer frame. The flaring region is characterized by a spherical comoving size $R' \sim \delta c t_{\rm var}$. The magnetic field is $B' \sim [2 \eta_B L_{\rm bol}/ (\delta^6 c^3 t_{\rm var}^2)]^{1/2}$, by setting $U'_B = \eta_B U'_{\rm rad}$, where $U'_{\rm B} = B'^2 / 8\pi$ is the comoving magnetic energy density and $U'_{\rm rad} = L'_{\rm bol} / 4 \pi R'^2 c$ is the comoving photon energy density of the flare. In the following, we set $\eta_B=1$. The ambient photon field is modeled by a broken power-law, with a break energy $\epsilon'_{\rm b}$ and spectral indices $a<2$ and $b>2$, respectively below and above $\epsilon'_{\rm b}$, such that ${\rm d}n'_\gamma/ {\rm d}\epsilon' = L_{\rm b}' / (4 \pi R'^2 c  {\epsilon}_{\rm b}'^2) \times ({\epsilon'}/{{\epsilon}_{\rm b}'})^{-x}$, where $x=a$ for $\epsilon'< {\epsilon}_{\rm b}' $ and $x=b$ for  $\epsilon'>{\epsilon}_{\rm b}' $, and $L'_{\rm b} \simeq L'_{\rm bol} / (1/(2-a)+1/(b-2))$. 

In this work, we consider that protons are present in the flaring region, and focus on high-energy neutrino flares produced though photohadronic interactions, namely photopion production $p \gamma \rightarrow N \pi$, where $N$ is a hadron and $\pi$ is a pion. Neutral pions decay into gamma rays $\pi^0 \rightarrow 2 \gamma$, whereas charged pions decay into leptons and neutrinos, according to the scheme $\pi^+ \rightarrow \mu^+ + \nu_\mu$, $\pi^- \rightarrow \mu^- + \bar{\nu}_\mu$, $\mu^+ \rightarrow e^+ + \bar{\nu}_\mu + \nu_e$ and $\mu^- \rightarrow e^- + \nu_\mu + \bar{\nu}_e$. The contribution of charged pions and muons is accounted for and the contribution of kaons is neglected. These particles usually contribute to the high-energy part of the neutrino spectrum with a lower contribution than pions and muons.
Each accelerated proton can produce typically $N_\pi=2\;{\rm to}\;5$ pions before losing a significant fraction of its initial energy (given the photopion production inelasticity and energy-loss timescale shown below), and about $50\%$ of the pions produced through photopion production are charged pions. We do not distinguish between neutrinos and anti-neutrinos. The multiplicity does not depend on energy or interaction channel, even if we note that more than one charged pion can be produced in the multi-pion production regime, for high photon energy in the parent particle rest frame. Finally, we do not account for neutrinos produced though neutron decay as they reach lower energies.
Moreover, we do not consider any hadronic background, as for flaring backgrounds, in most of the cases it leads to subdominant contributions to the high-energy neutrino production \citep[see related discussion in][]{Guepin17}. Steady hadronic background can be dominant in several categories of explosive transients associated with the death of massive stars, therefore these cases should be considered with care by evaluating the photon and baryon densities and comparing the photohadronic and purely hadronic interaction timescales. 

Protons can be accelerated and lose energy before they escape the region, or interact with the ambient photon field and produce pions. The secondary particles produced though photopion production can also experience acceleration and energy losses before they decay and produce neutrinos. We adopt a simplified approach, where the energy evolution of particles is described by the generic equation
\begin{equation}\label{Eq:energy}
\frac{\d E'}{\d t'} = \left.\frac{\d E'}{\d t'}\right|_{\rm gain} - \left.\frac{\d E'}{\d t'}\right|_{\rm loss} \, .
\end{equation}
The gain and loss terms correspond respectively to acceleration and energy loss processes. They can be estimated using characteristic timescales $t_{\rm acc}$ for the acceleration process, and $t_{\rm loss}$ for each energy-loss process considered. In addition to this treatment of quasi-continuous particle energy evolution, we adopt a random treatment for photopion production, charged pion and muon decay, and escape from acceleration process. In the following, we detail the various timescales and random treatments mentioned above.

A large variety of physical mechanisms leading to particle acceleration have been explored in the literature, such as shock acceleration, turbulent acceleration, shear acceleration, unipolar induction or magnetic reconnection, as a few examples. In this work, we adopt a general phenomenological description of particle energization, described by the timescale $t'_{\rm acc} \sim E'/\dot{E'}_{\rm gain}$. With astrophysical plasmas being highly conductive, the typical acceleration timescale can be related to the Larmor time $t'_{\rm acc} \sim \eta_{\rm acc}^{-1} t'_{\rm L} = \eta_{\rm acc}^{-1} E'/ c e B'$, with $\eta_{\rm acc}\leq1$. The case $\eta_{\rm acc}=1$ corresponds to a maximally efficient acceleration, and can be obtained for instance in the case of linear acceleration in perfectly conducting and relativistic plasmas. Physical mechanisms leading to particle acceleration often involve scattering against magnetic inhomogeneities. In this specific context, $t'_{\rm acc} \propto t'_{\rm L} (c t'_{\rm L}/l_B)^{\alpha_{\rm t}}$ \citep[e.g.][]{Lemoine09}, where $l_B$ is the coherence length of the magnetic field, and  $\alpha_{\rm t}=1$ for $c t'_{\rm L} > l_B$ and $\alpha_{\rm t}=\beta_{\rm t}-1$ for $c t'_{\rm L} < l_B$, with $\beta_{\rm t}$ the spectral index of the turbulence power spectrum. For a given energy, $t'_{\rm acc}$ reaches its minimum value for $c t'_{\rm L} = l_B$, which corresponds to the so-called Bohm regime $t'_{\rm acc} \propto t'_{\rm L}$. In the following, we consider a maximally efficient acceleration with $t'_{\rm acc} = t'_{\rm L}$, and thus study the conditions for which secondary acceleration should have the strongest impact. In addition to this description of acceleration timescale, we consider that the particles experience some kind of scattering process and have a probability $p_{\rm esc}$ of escaping the acceleration process every scattering or cycle---typically every Larmor time $t'_{\rm L}$ in the regime studied. This classical description of stochastic collisions, with a constant escape probability $p_{\rm esc}$ and a constant fractional energy increase $\eta_E \equiv {\rm d}E/E$ per cycle, produces power-law particle spectra ${\rm d}N/{\rm d}E \propto E^{-\alpha}$ with $\alpha = 1 - \log (1-p_{\rm esc})/\log (1+\eta_E) $. It allows to treat self-consistently the acceleration, energy-losses and interactions of protons, charged pions and charged muons.

As regards energy-loss processes, we consider synchrotron and adiabatic energy losses, for which $t'_{\rm syn} = 6\pi (m c^2)^2 / [(m_e/m)^2 \sigma_{\rm T} c B'^2 E' ]$ and $t'_{\rm dyn} = \delta t_{\rm var}$. Inverse Compton and Bethe-Heitler processes, that are often subdominant over the energy range of interest, are neglected. Photohadronic interactions of pions, that can lead to pion cascades, are also neglected. We discuss their impact in appendix~\ref{app:photohadronic_secondaries}. 

The aforementioned simplifications allow to solve equation~\ref{Eq:energy} analytically, under a quasi-continuous approximation, and obtain an indicative time evolution of the energy of the particles. Equation~\ref{Eq:energy} becomes
\begin{align}\label{Eq:equadiff}
\dfrac{\d {E'}}{\d t'} &= - A_2 \, {E'}^2 - A_1 \, {E'} + A_0 \, , \nonumber \\
&=  \frac{\Delta}{4 A_2}  \left[ 1 - \frac{1}{\Delta} \left(2 A_2 \, {E'}+ A_1 \right)^2 \right]\, ,
\end{align}
with:
\begin{align}\label{Eq:params}
\begin{cases} 
A_2 &= \left( m_{\rm e}/m \right)^2 \sigma_{\rm T} c {B'}^2 / [6\pi (m c^2)^2]  \, ,\\
A_1 &= (\delta \, t_{\rm var})^{-1} \, , \\
A_0 &= c e B' \, ,\\
\Delta &= A_1^2 + 4 A_0 A_2 \, .
\end{cases}
\end{align}
We set $x^2 =  \left(2 A_2 \, {E'} + A_1 \right)^2 / \Delta$ and $x_0 = x(t'=0)$. For $x=1$, $E' \equiv E'_{\rm lim} = ( \sqrt{\Delta} - A_1 )/2A_2$. For $E'(t'=0) < E'_{\rm lim}$, the energy increases with time
\begin{equation}
E'(t') = \frac{ \sqrt{\Delta}}{2 A_2} \, {\rm th} \left( \frac{1}{2} \sqrt{\Delta} \, t' + {\rm argth} (x_0) \right) - \frac{A_1}{2 A_2}  \, ,
\end{equation} 
and for $E'(t'=0) > E'_{\rm lim}$, the energy decreases  
\begin{equation}\label{Eq:decrease}
E'(t') = \frac{\sqrt{\Delta}}{2 A_2} \, {\rm coth} \left( \frac{1}{2} \sqrt{\Delta} \, t' + {\rm argcoth} (x_0) \right) - \frac{A_1}{2 A_2}  \, .
\end{equation}

Equations~\ref{Eq:equadiff}, \ref{Eq:params} and \ref{Eq:decrease} are still valid without acceleration, with $A_0 \equiv 0$. In the case of charged pions and muons, their energies evolve until their decay. We assume that the times at which photopion production, charged pion decay or charged muon decay occur are characterized by exponential distributions, with characteristic times $t'_{p\gamma}$, $ t'_\pi =  \tau_\pi  E'_{\pi} / m_\pi c^2 $ and $ t'_\mu = \tau_\mu E'_{\mu} / m_\mu c^2 $. The energy-loss timescale related to photopion production 
\begin{equation}
t_{p\gamma}'^{-1} \simeq \frac{\left\langle \sigma_{p\gamma} \kappa_{p\gamma} \right\rangle L_{\rm bol} }{ 4 \pi \delta^5 c^2 t_{\rm var}^2 \epsilon_{\rm b}}\frac{1}{1-a} \left[\frac{a-b}{1-b}-\left(\frac{\epsilon_{\rm th}}{{\epsilon}_{\rm b}} \right)^{1-a}\right] \, ,
\end{equation}
depends on the photon spectrum, the cross section and inelasticity of photopion production $\sigma_{p\gamma}$ and $\kappa_{p\gamma}$, and the interaction threshold energy $\epsilon_{\rm th}$. 

Combining this analytical energy evolution and random treatment of photopion production, pion and muon decay and acceleration allows to explore efficiently the properties of high-energy neutrino flares in the parameter space of transient sources. The maximum neutrino energies described in section~\ref{sec:Emax} are obtained by computing only the maximum energies of protons, charged pions and muons. The neutrino spectra presented in section~\ref{sec:Spec} require to compute the energy evolution in various energy bins. We initially inject a mono-energetic spectrum of protons in the flaring region, with a Lorentz factor $\gamma_p=1$ and follow the energy evolution of a large number of particles. All particles entering or produced in the flaring region are accelerated, with a constant probability $p_{\rm esc}$ of escaping the acceleration process at each scattering time. After they escape the acceleration process, they only lose energy before they interact, decay or escape the flaring region. The number of particles in each energy bin pondered with the adequate normalization gives the final spectra. This normalization depends on the source properties and on the primary spectra---the mono-energetic spectrum for protons, the spectrum of protons undergoing photopion production for charged pions and the spectrum of decayed charged pions for charged muons.  In addition to the semi-analytical calculation of high-energy neutrino spectra described above, we use a general propagation and interaction code, comprised of modules from \textsc{CRPropa} \citep{CRPropa_v3.1} and a Monte Carlo code \citep[e.g.][]{Kotera09}, in which we implement our phenomenological treatment of particle energization for protons, charged pions and charged muons. This code gives similar high-energy neutrino spectra, and allows to compute precisely the spectrum of escaping protons. In addition to tables of interaction or energy-loss lengths of the different processes described above, we also include in this code subdominant processes, as inverse Compton and Bethe-Heitler losses.

Finally, we note that protons escaping the flaring region do not contribute to the high-energy neutrino spectrum. The neutrinos produced during the large-scale propagation of protons from the source to the Earth do not contribute to the neutrino flare, in the absence of spatial and temporal coincidence due to proton deflections during their propagation. We note that charged pions and muons escaping the source before they decay could still contribute to the neutrino flare, due to their short decay times. The high-energy neutrino spectra described in section~\ref{sec:Spec} are not impacted by the escape of charged pions and muons, as acceleration, synchrotron and decay are the dominant processes.

\section{Maximum neutrino energy}
\label{sec:Emax}

\begin{figure*}[!tp]
\centering
{\includegraphics[width=0.95\textwidth]{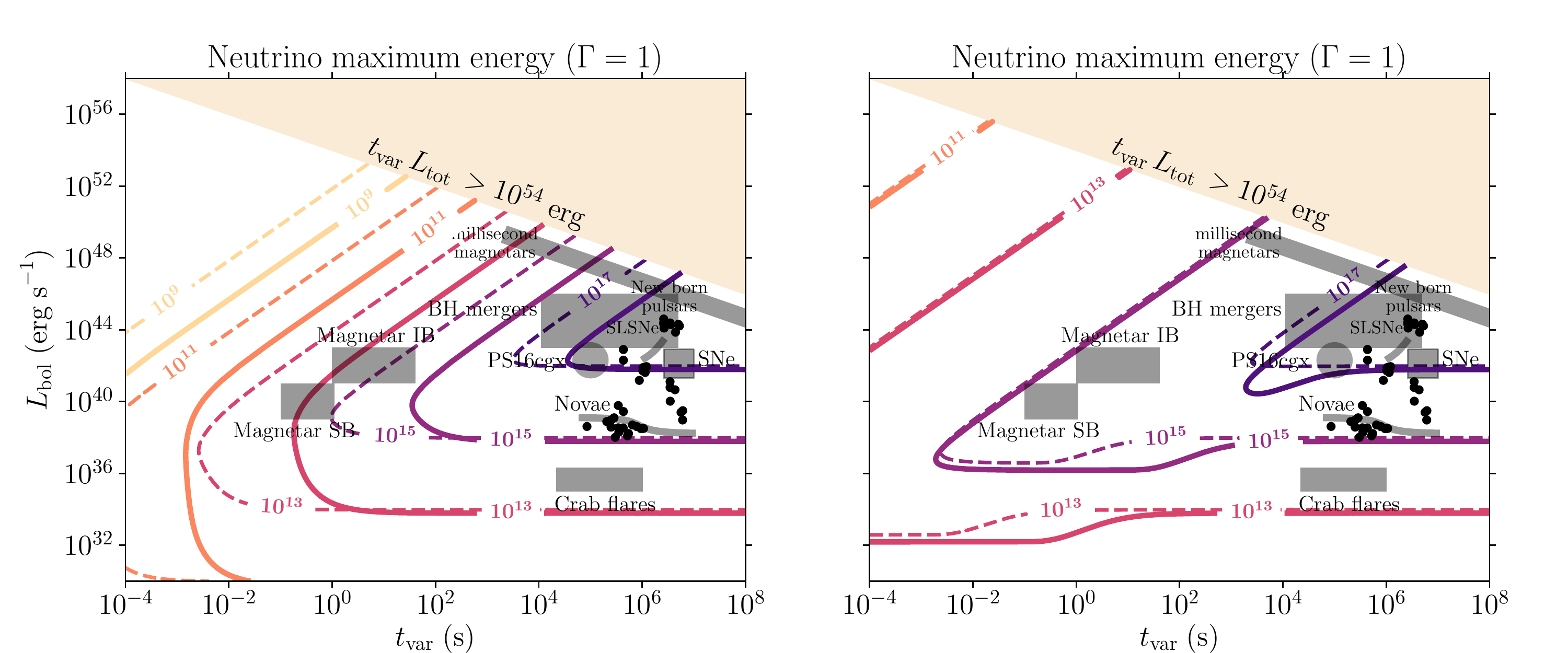}}
{\includegraphics[width=0.95\textwidth]{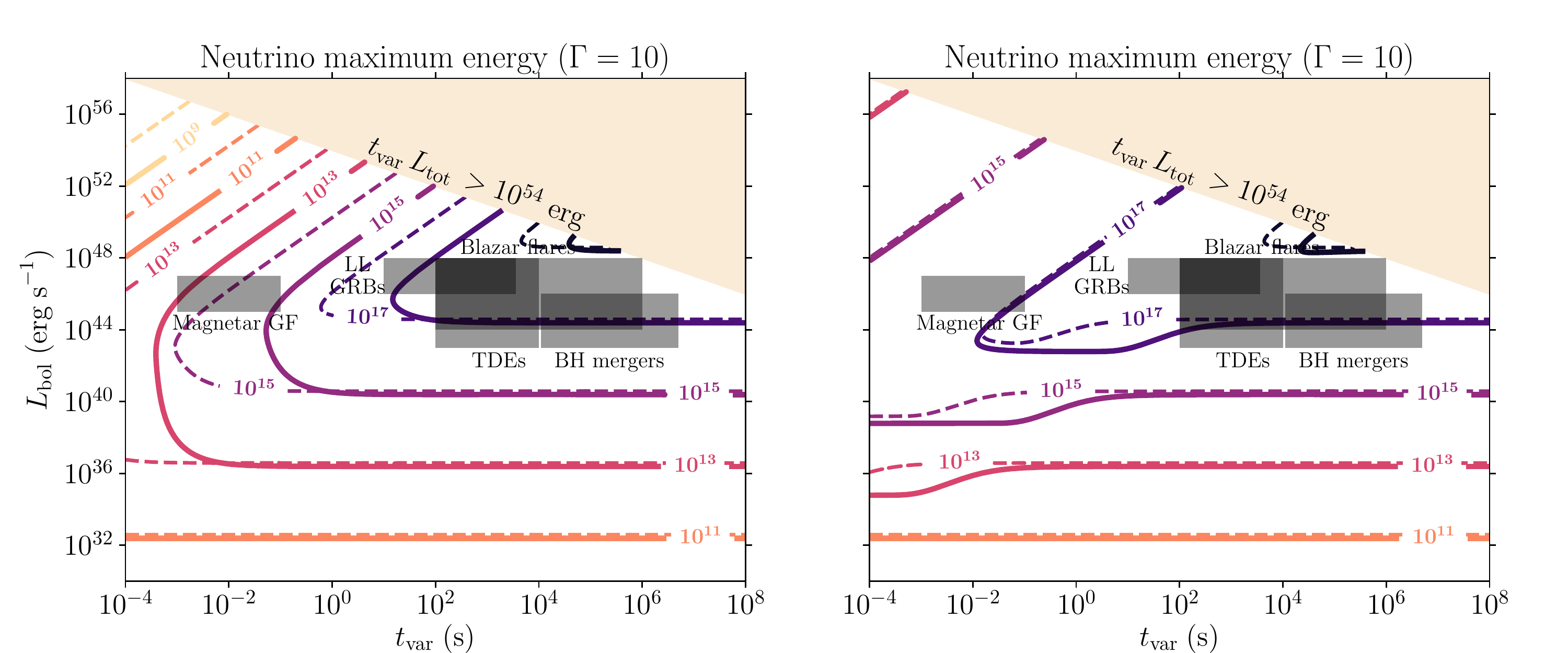}}
{\includegraphics[width=0.95\textwidth]{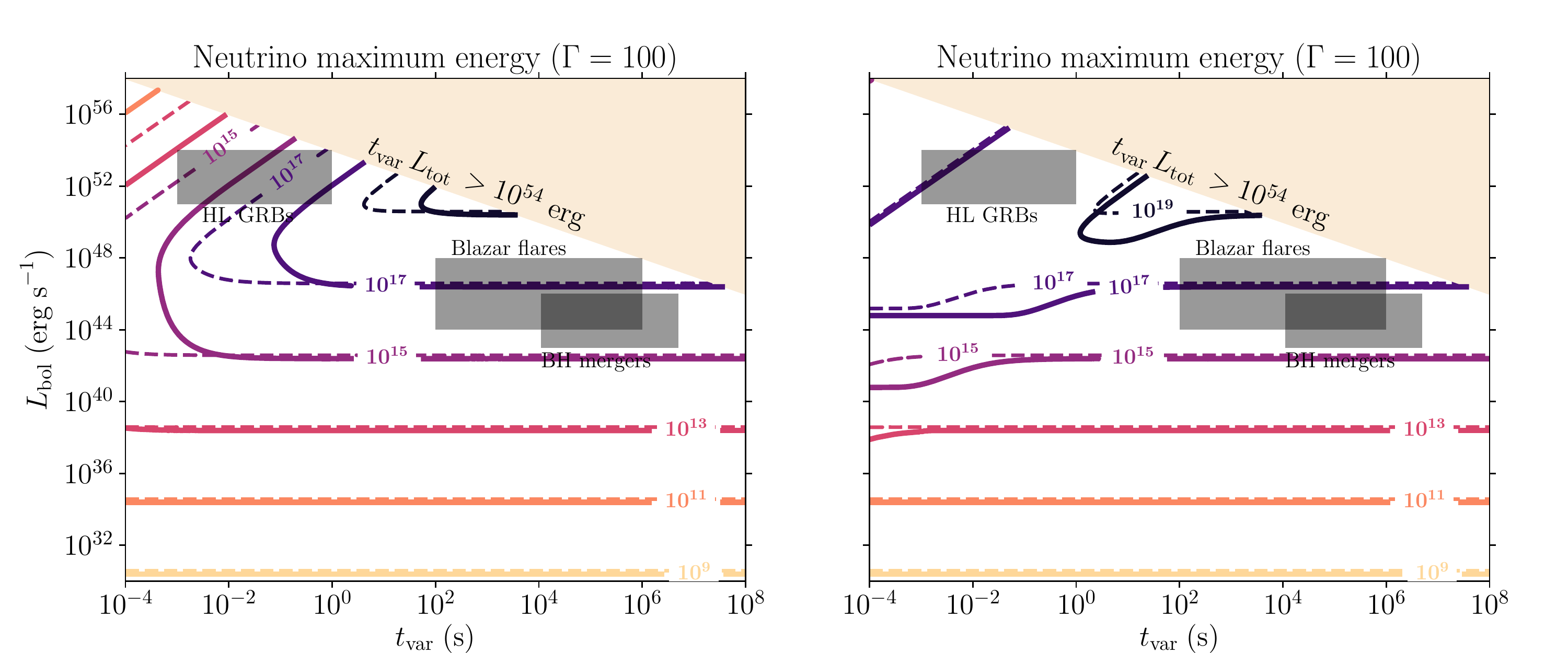}}
\caption{Maximum neutrino energy without (left column) and with (right column) secondary acceleration, as a function of the variability timescale $t_{\rm var}$ and the bolometric luminosity $L_{\rm bol}$ of a flaring source, with bulk Lorentz factor $\Gamma = 1, 10, 100$ (from top to bottom). Dashed and solid contours indicate the maximum energy of neutrinos produced respectively through charged pion and charged muon decay. Overlayed are examples of location of benchmark explosive transients in the $t_{\rm var}-L_{\rm bol}$ parameter-space. The pale orange region indicates the domain where no source is expected to be found due to the excessive energy budget.}\label{fig:Enumax}
\end{figure*}

The maximum energy of high-energy neutrinos produced through photohadronic interactions can be represented in parameter space $t_{\rm var}$, $L_{\rm bol}$, for different bulk Lorentz factors ($\Gamma=1,10$ and $100$), as illustrated in figure~\ref{fig:Enumax}. We show both the maximum energy of neutrinos produced through charged pion and muon decays, respectively with dashed and solid contours, and compare the energies obtained without or with secondary acceleration, in the left and right columns. In this calculation, we assume that the fractions of parent energies deposited in daughter species are $\chi_{p \rightarrow \pi^\pm} \simeq 0.2$, $\chi_{\pi^\pm \rightarrow \mu^\pm} \simeq 0.8$, $\chi_{\pi^\pm \rightarrow \nu_{\rm direct}} \simeq 0.2$ and $\chi_{\mu^\pm \rightarrow \nu_{\rm direct}} \simeq 0.3$, from the kinematics of photomeson production \citep[see e.g.][]{Hummer10}, and from the kinematics of two and three body decays. The notation $\chi_{\pi^\pm \rightarrow \nu_{\rm direct}}$ refers to the muon neutrinos directly produced through charged pion decay $\pi^+ \rightarrow \mu^+ + \nu_\mu$ or $\pi^- \rightarrow \mu^- + \bar{\nu}_\mu$, whereas the notation $\chi_{\mu^\pm \rightarrow \nu_{\rm direct}}$ refers to the muon and electron neutrinos directly produced through charged muon decay $\mu^+ \rightarrow e^+ + \bar{\nu}_\mu + \nu_e$ and $\mu^- \rightarrow e^- + \nu_\mu + \bar{\nu}_e$.

We also locate typical flaring source categories in the parameter space: magnetar short bursts (SB), intermediate bursts (IB) and giant flares (GF), Crab flares, novae, thermonuclear and core-collapse supernovae (SNe), super-luminous supernovae (SLSNe), black hole (BH) mergers, low-luminosity (LL) and high-luminosity (HL) gamma-ray bursts (GRBs), blazar flares, and tidal disruption events (TDEs). For more detail about these source categories and their location in parameter space, see \cite{Guepin17}. A grey band has been added to locate pulsars, in particular millisecond magnetars and new-born pulsars with millisecond periods, in the parameter space with $\Gamma=1$. The contribution of these sources to observed cosmic-ray and neutrino fluxes have been studied for instance in \cite{Venkatesan97, Blasi00, Arons03, Fang12, Fang13, Lemoine15, Kotera15}. We locate these sources in the parameter space by associating the bolometric luminosity to the spin-down luminosity and the variability time-scale to the spin-down time-scale, respectively $L_{\rm sd} = c B_{\rm p}^2 R^6 / [4 (c P/2\pi)^4]$ and $t_{\rm sd} = 9 I c^3 P^2 / 8 \pi^2 B_{\rm p}^2 R^6$, where $B_{\rm p}$, $R$, $P$ and $I$ are the polar magnetic field, radius, initial spin period and moment of inertia or the pulsar.

The left column of figure~\ref{fig:Enumax} shows neutrino maximum energy without the impact of secondary acceleration. We note that neutrinos produced through charged pion and muon decays do not necessarily reach the same energies, especially when important secondary energy-losses are at play. At high bolometric luminosities (e.g. $L_{\rm bol} > 10^{42}$\,erg\,s$^{-1}$ for $\Gamma=1$) synchrotron losses are dominant due to the large magnetic fields. At lower luminosities and small variability timescales, where the contour lines are vertical, adiabatic losses are dominant. In these regions of the parameter space, energy losses occur before the decay of the charged pions or muons. Due to the larger disintegration time of charged muons, for the same parameters $t_{\rm var}$, $L_{\rm bol}$ and $\Gamma$, they lose more energy than charged muons before decaying. 

The right column of figure~\ref{fig:Enumax} illustrates the influence of secondary acceleration. It is only effective in a portion of the parameter space, where $t'_{\rm acc} < t'_{\rm dec}$. This gives the constraint $L_{\rm bol} > (m^2 c^5 \delta^6 / 2 \tau^2 e^2 \eta_B) t_{\rm var}^2 $ for efficient secondary acceleration, where $m$ and $\tau$ are the mass and decay time of charged pions or muons. Due to the larger decay time of muons, their acceleration can occur at lower $L_{\rm bol}$ for fixed $t_{\rm var}$. Consequently, neutrinos produced through muon decay can reach significantly higher energies than neutrinos produced through pion decay in the transition region where secondary acceleration becomes efficient. In the rest of the parameter space impacted by secondary acceleration, neutrinos produced by the decay of charged pion and muon reach similar energies. From this parameter space study, we see that many objects could be influenced by secondary acceleration: millisecond magnetars, magnetar intermediate bursts, short bursts and magnetar giant flares, low-luminosity gamma-ray bursts, high-luminosity gamma-ray bursts, and tidal disruption events and blazar flares to a slightly lesser extent.

We emphasize that the maximum energies illustrated in these parameter spaces are indicative. First, the factor $\eta_B=1$ which sets the magnetic field influences acceleration but also synchrotron losses. Lower $\eta_B$ lead to less efficient acceleration thus lower neutrino energy where decay is dominant, but lower synchrotron losses thus higher neutrino energy where synchrotron is dominant. We note that this simple example does not account for additional energy losses such as photopion interactions, see appendix~\ref{app:photohadronic_secondaries}. Second, in the case of secondary acceleration, only a small fraction of secondaries could be accelerated and thus only a small fraction of neutrinos could be produced at this maximum energy. It is therefore important to determine the peak energy of the neutrino spectrum, and case-by-case studies are required. In the following, we consider several fiducial examples or sources classes, in order to illustrate the consequences of secondary acceleration in different regions of the parameter space of explosive transients. 

\section{High-energy neutrino spectra}
\label{sec:Spec}

As described in section~\ref{sec:model}, a mono-energetic spectrum of protons is injected in the flaring region, and protons undergo acceleration and energy-loss processes before escaping the flaring region or producing a pion. The slope $\alpha$ of the proton spectrum escaping the acceleration process ${\rm d}N_p/{\rm d}E_p \propto E_p^{-\alpha}$ is determined by the probability $p_{\rm esc}$ of escaping the acceleration region. This probability also controls the efficiency of secondary acceleration. In the following we compare two cases:  $\alpha \simeq 1.1$, obtained for $p_{\rm esc,low} \simeq 0.067$ and $\alpha \simeq 1.9$, obtained for $p_{\rm esc,high} \simeq 0.46$. For $\alpha=1.1$, most of the cosmic-ray energy is channeled into the highest energies, whereas for $\alpha=1.9$, the energy is more spread across the spectrum. The high-energy cut-off of the proton spectrum is determined by the competition between acceleration and energy losses. 

In order to normalize the spectra, we consider $\mathcal{E}'_{\rm acc, p} = \eta_{\rm p} \,\mathcal{E}'_{\rm rad}$, where $\mathcal{E}'_{\rm acc, p}$ is the energy required to accelerate protons in the comoving frame,   $\mathcal{E}'_{\rm rad} \sim \delta^{-3} L_{\rm bol} t_{\rm dur}$ is the comoving non-thermal energy radiated during the total duration of the emission $t_{\rm dur}$ and $\eta_{\rm p}$ is a constant. We note that the factor $\eta_{\rm p}$ depends on the radiation efficiency and the baryon loading, and can vary by many orders of magnitude. Considering the total energy for acceleration $\mathcal{E}'_{\rm acc}$ and the energy required to accelerate electrons (and positrons) $\mathcal{E}'_{\rm acc,e}$ we have $\mathcal{E}'_{\rm acc} = \mathcal{E}'_{\rm acc,p} + \mathcal{E}'_{\rm acc,e}$, with $\mathcal{E}'_{\rm acc,p} = \eta_{\rm acc, p} \, \mathcal{E}'_{\rm acc}$ and  $\mathcal{E}'_{\rm acc,e} = \eta_{\rm acc, e} \, \mathcal{E}'_{\rm acc}$ such that $\eta_{\rm acc, p} + \eta_{\rm acc, e} =1$. Moreover, the accelerated protons and electrons radiate a fraction of their acceleration energy, for instance through synchrotron radiation for protons and electrons or $\pi^0$ production for protons, such that $\mathcal{E}'_{\rm rad,p}=\eta_{\rm rad, p} \, \mathcal{E}'_{\rm acc,p}$ and $\mathcal{E}'_{\rm rad,e}=\eta_{\rm rad, e} \, \mathcal{E}'_{\rm acc,e}$. With this simple description, we obtain $\mathcal{E}'_{\rm rad} = \mathcal{E}'_{\rm rad,p} +  \mathcal{E}'_{\rm rad,e} = \mathcal{E}'_{\rm acc,p} (\eta_{\rm rad, p}+\eta_{\rm rad, e} \eta_{\rm acc, p}/\eta_{\rm acc, e})$ and $\eta_{\rm p} = [\eta_{\rm rad, p}+\eta_{\rm rad, e}\, \eta_{\rm acc, p}/(1-\eta_{\rm acc, p})]^{-1}$. Therefore, if both electrons and protons radiate most of their acceleration energy, $\eta_{\rm p} \sim \eta_{\rm acc, p} < 1$ and the normalization can be directly constrained by the baryon loading. However, if only a small fraction of the acceleration energy of protons or electrons is radiated, for instance if radiation is not limiting for acceleration or the spectra are soft, or if a part of the non-thermal photons interact and produce delayed emissions, we can have $\eta_{\rm p} \gg 1$. We note that in our case studies, the maximum energies of protons and electrons are mostly limited by radiation, through synchrotron or $\pi^0$ production. Moreover, the escape probabilities $p_{\rm esc}$ considered lead to hard spectra. Therefore, if protons and electrons experience the same acceleration processes, most of the acceleration energy is radiated, and in this case $\eta_{\rm p}=1$ can be considered as an upper bound. To obtain the fluence on Earth, we also include a factor $1/(4\pi d_{\rm L}^2)$, where $d_{\rm L}$ is the luminosity distance from the source, and a factor $\delta^3$, which accounts for the transformation from the comoving frame to the observer frame.

\subsection{Case studies}

Our case studies focus on several regions of the parameter space of explosive transients that can be related to specific source categories. From section~\ref{sec:Emax}, different types of transient emissions from highly magnetized pulsars (also magnetars) can be affected by secondary acceleration. As mentioned earlier, magnetars have been identified in many studies as promising candidates for the acceleration of cosmic rays and the production of secondary high-energy neutrinos, for instance \cite{Blasi00, Fang12, Fang13, Lemoine15, Kotera15}. The case of newborn magnetars with millisecond periods illustrates a non-relativistic source class, whereas the case of magnetar giant flares involves relativistic outflows. Moreover, tidal disruption events, low-luminosity gamma-ray bursts and blazar flares are examples of relativistic outflows, whose properties partially overlap in the parameter space of explosive transients. In these overlapping regions, they can be similarly affected by secondary acceleration. Therefore, we choose to describe the case of jetted tidal disruptions, while keeping in mind that this case study can be used as a benchmark example for low-luminosity gamma-ray bursts and blazar flares. We note that beyond standard scenarios involving gamma-ray bursts \citep[e.g.][]{Waxman97_GRB,Murase06_flares,Murase08,Meszaros15} and active galactic nuclei \citep[e.g.][]{Bednarek99,Atoyan01,Halzen05,Dermer14,Petropoulou16,Murase18,Gao19}, jetted tidal disruptions have also been proposed as candidate sources for the production of high-energy cosmic rays and neutrinos \citealp{Wang11, Senno16b, Dai16, Lunardini16, Wang16, Zhang17, Biehl17, Guepin18b}. These case studies are associated with different types of photon fields, that we simply model by hard or soft broken power laws, and we can thus assess their impact on the high-energy neutrino spectrum. From section~\ref{sec:Emax}, all these source categories are affected by strong secondary synchrotron losses and should be affected differently by secondary acceleration.

First, we focus on the case of millisecond magnetars formed from binary neutron star mergers, as magnetars formed in core-collapse supernovae are surrounded by a massive envelope, and thus their high-energy neutrino production is dominated by purely hadronic interactions. We follow the approach of \cite{Fang17}. As the magnetic field at the pole of the neutron star has a strong influence on the spin-down luminosity and thus the associated bolometric luminosity, we consider two cases $B_{\rm p}=10^{14}\,{\rm G}$ and $B_{\rm p}=10^{15}\,{\rm G}$. For both cases, we have $\Gamma=1$, $t_{\rm dur}=t_{\rm var}$ and we consider the spin-down luminosity at $t=t_{\rm var}$, such as $L_{\rm sd} \propto (1+t/t_{\rm sd})^{-2}$. For $B_{\rm p}=10^{14}\,{\rm G}$, we have $t_{\rm var} = 3\times 10^5\,{\rm s}$ and $L_{\rm bol} = 4\times 10^{46}\,{\rm erg\,s}^{-1}$, and for $B_{\rm p}=10^{15}\,{\rm G}$, $t_{\rm var} = 3\times 10^3\,{\rm s}$ and $L_{\rm bol} = 4\times 10^{48}\,{\rm erg\,s}^{-1}$. In both cases, the system should be surrounded by hadronic material, typically $M_{\rm ej} = 10^{-2} M_\odot$. In the last case, purely hadronic interactions play a minor role at the lowest energies and can thus be neglected; nevertheless, we have included them in our neutrino spectra for sake of consistency. From \cite{Fang17}, optical/UV/X-ray thermal and non-thermal radiation backgrounds should contribute to the production of high-energy neutrinos. For simplicity and to comply with our model, the target for photohadronic interactions is modeled as a broken power law with a hard spectral index $a = -1$ below $\epsilon_{\rm b} = 1\,{\rm eV}$, accounting for the thermal contribution, and $b = 3.1$ above, accounting for the non-thermal tail.

Second, we study the case of magnetar giant flares. In our model, they are characterized by a variability timescale $t_{\rm var}=10^{-2}\,{\rm s}$, a bolometric luminosity $L_{\rm bol} = 2\times 10^{47}\,{\rm erg\,s}^{-1}$, a bulk Lorentz factor $\Gamma=10$ and a total duration $t_{\rm dur} = 1\,{\rm s}$. The photon spectrum used as a target for photohadronic interactions is a power law characterized by a hard spectral index $a = 0.1$ below $\epsilon'_{\rm b} = 5\,{\rm keV}$ and $b = 3.1$ above. 

Third, we describe tidal disruptions powering relativistic jets by a variability timescale $t_{\rm var}=10^2\,{\rm s}$, a bolometric luminosity $L_{\rm bol} = 10^{48}\,{\rm erg\,s}^{-1}$, a bulk Lorentz factor $\Gamma=10$ and a total duration $t_{\rm dur} = 10^5\,{\rm s}$. The photon spectrum used as a target for the interactions is characterized by a break energy $\epsilon'_{\rm b} = 0.5\,{\rm keV}$, a soft spectral index below the break $a=1.8$ and a spectral index above the break $b = 3.1$. Due to the scarcity of observations of jetted tidal disruptions, the photon spectrum is not well constrained, thus we consider a soft spectrum, that can be adapted to the cases of low-luminosity gamma-ray bursts and blazar flares.

The length scales of the different energy-gain and energy-loss processes accounted for in our calculation are illustrated in appendix~\ref{app:MFP} for protons, for the case studies described above. The maximum energies of accelerated particles, and more generally the energies at which the different processes occur, can be related to the proton and neutrino spectra shown in the following.

\subsection{Proton spectrum and secondary losses}

\begin{figure*}[htpb]
\centering
\includegraphics[width=0.95\textwidth]{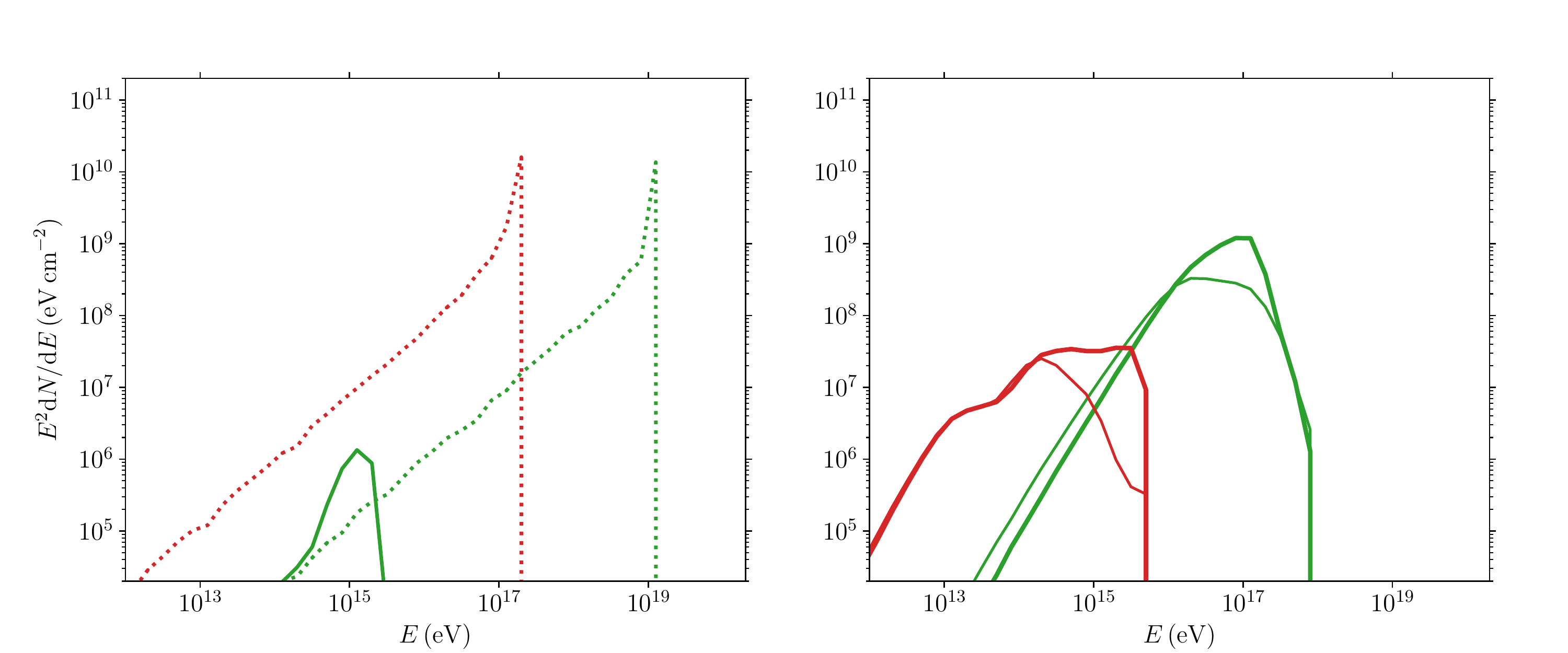}
\includegraphics[width=0.95\textwidth]{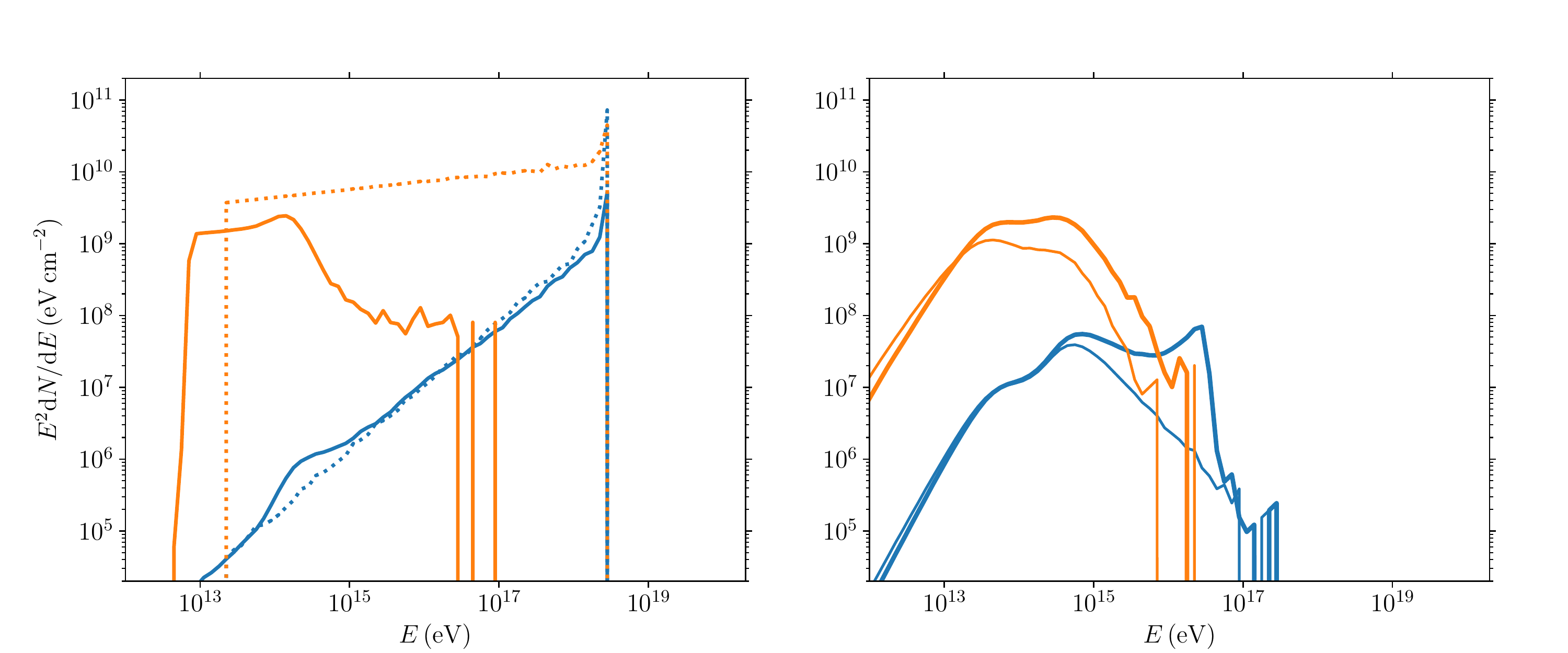}
\includegraphics[width=0.95\textwidth]{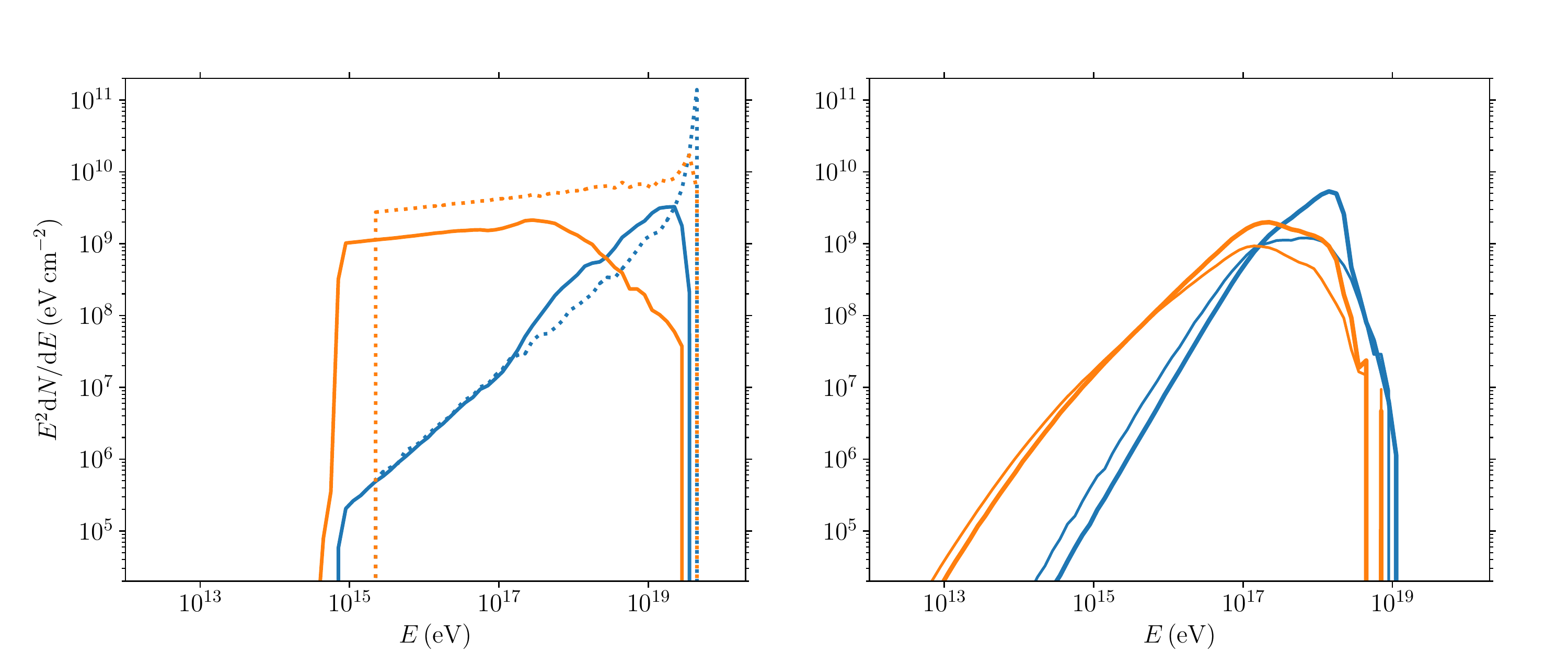}
\caption{Proton and neutrino spectra for millisecond magnetars (top), magnetar giant flares (middle) and jetted tidal disruptions (bottom). For millisecond magnetars, we compare $B_{\rm p}=10^{14}\,{\rm G}$ and $B_{\rm p}=10^{15}\,{\rm G}$ (green and red lines) for $p_{\rm esc,low}$. For magnetar giant flares and jetted tidal disruptions, we compare $p_{\rm esc,low}$ and $p_{\rm esc,high}$ (blue and orange lines). Left column: spectra of purely accelerated protons escaping the acceleration process without photohadronic interactions (dotted lines) and protons escaping the flaring region (solid lines). Right column: neutrino spectra without and with secondary acceleration (thin and thick lines). }\label{fig:spec_cases}
\end{figure*}

The proton and neutrino spectra that we obtain for these three case studies are illustrated in figure~\ref{fig:spec_cases}. The indicative luminosity distances chosen are respectively $d_{\rm L}=100\,{\rm Mpc}$, $d_{\rm L}=100\,{\rm kpc}$ and $d_{\rm L}=100\,{\rm Mpc}$. For each case, we compare two proton spectra: the spectrum obtained only with acceleration and continuous energy losses when protons just escape acceleration region, and the spectrum of protons escaping the source. These spectra are thus illustrative only and do not correspond to any observable spectrum. In particular, the spectrum of protons escaping the source does not account for propagation and interaction effects outside of the source. In order to obtain a reasonable statistics at the highest energies, especially for $p_{\rm esc,high}$, for which the number of particles decrease with the energy, the low energy part of the proton spectra are not calculated. At these energies the high-energy neutrino production is inefficient, thus it does not impact our results. For each case study, we also compare two all-flavor neutrino spectra, obtained without and with secondary acceleration. These spectra do not account for the neutrino adiabatic losses due to the universe expansion as we focus on the impact of secondary acceleration.

In this work, two acceleration efficiencies are compared, with $p_{\rm esc,low}$ and $p_{\rm esc,high}$, in the so-called Bohm regime. This model produces power-law with indices $\alpha \simeq 1.1$ and $\alpha \simeq 1.9$. For millisecond magnetars, we focus on the case $p_{\rm esc,low}$, given the typical mono-energetic injection of cosmic rays integrated over spin-down time that produces a hard cosmic-ray spectrum, but we compare the cases $B_{\rm p}=10^{14}\,{\rm G}$ and $B_{\rm p}=10^{15}\,{\rm G}$. For magnetar giant flares and jetted tidal disruptions, we compare $p_{\rm esc,low}$ and $p_{\rm esc,high}$. We note the sharp peak of the accelerated proton spectra that appears for $p_{\rm esc,low}$, which is characteristic of the maximally efficient acceleration process considered, as a large amount of injected particles pile up at the energy at which energy losses become dominant. When synchrotron radiation is the limiting mechanism for acceleration, the spectrum of escaping protons also shows a sharp peak at the maximum energy, which is the case for the magnetar giant flare case study with $p_{\rm esc,low}$. When photomeson production is the energy-loss mechanism limiting proton acceleration, this peak is less pronounced due to our modeling of photomeson production as a random process. In this case, protons interact before being accelerated to the highest energies allowed by the competition between acceleration and synchrotron or adiabatic losses. We note that this effect can be more pronounced for soft photon spectra than for hard spectra, as in this case the mean free path of photopion production decreases with increasing energy. As expected, photomeson production produces a break or a cut-off in the spectra of escaping protons. For efficient photomeson production, only a small fraction of the highest-energy protons escape the source. At these energies, the slope of the spectrum of escaping protons $\alpha_{\rm esc}$ is determined by the accelerated proton and photon spectra, such that $\alpha_{\rm esc}=\alpha$ for a hard photon spectrum, as illustrated by the case of case of magnetar giant flares.

Regardless whether or not the secondary acceleration if accounted for, $p_{\rm esc}$ can have an important impact on the peak neutrino flux, as it influences the slope of the proton spectrum. The high-energy neutrino peak flux can be more than one order of magnitude lower for $p_{\rm esc,low}$ than for $p_{\rm esc,high}$, as shown in figure~\ref{fig:spec_cases}, as the number of particles piling up at the peak energy due to secondary losses is lower than the flux decrease due to the spectral index difference. These effects can be confirmed by comparing estimates of the cumulated proton fluxes at the energy corresponding to the neutrino peak fluxes $E_{\rm pk}^2 \Phi(E_{\rm pk})+E_{\rm pk} \int_{E_{\rm pk}}^{E_{\rm max}} {\rm d}E \Phi(E)$.

The peak energies of the neutrino spectra calculated in our case studies without secondary acceleration are consistent with the maximum energies calculated in section~\ref{sec:Emax}. Moreover, we note that the peaks of the $\nu_{\pi^\pm \rightarrow \nu_{\rm direct}}$ and $\nu_{\mu^\pm \rightarrow \nu_{\rm direct}}$ neutrino spectra can be distinguished. With our assumptions, $\nu_{\pi^\pm \rightarrow \nu_{\rm direct}}$ are $100\%$ of muon neutrinos and $\nu_{\mu^\pm \rightarrow \nu_{\rm direct}}$ are $50\%$ of muon neutrinos and $50\%$ of electron neutrinos. After propagation and oscillation, the flavor ratios of transient neutrino flares detected at Earth should therefore be energy dependent \citep[e.g.][]{Bustamante19}, which could be probed by next-generation neutrino detectors. Despite the fact that more $\nu_{\mu^\pm \rightarrow \nu_{\rm direct}}$ are produced, the peak of the $\nu_{\pi^\pm \rightarrow \nu_{\rm direct}}$ neutrino spectra can be higher than the peak of the  $\nu_{\mu^\pm \rightarrow \nu_{\rm direct}}$ neutrino spectra, due to the gap between $\nu_{\pi^\pm \rightarrow \nu_{\rm direct}}$ and $\nu_{\mu^\pm \rightarrow \nu_{\rm direct}}$ energies, in the case of large secondary energy losses. 

As shown in figure~\ref{fig:Enumax}, our case studies are differently impacted by secondary losses, which affects the neutrino spectra by shifting the maximum neutrino energies below the typical $5\%$ of the maximum proton energies obtained without secondary losses. For all case studies, synchrotron losses have an impact on the proton, charged pion and charged muon spectra. As illustrated by the energies of the proton spectral peaks in figure~\ref{fig:spec_cases}, they affect millisecond magnetars with $B_{\rm p}=10^{14}\,{\rm G}$ less than the ones with $B_{\rm p}=10^{15}\,{\rm G}$. Among our case studies, millisecond magnetars with $B_{\rm p}=10^{15}\,{\rm G}$ and magnetar giant flares are the most impacted by secondary synchrotron losses. The effect is less pronounced for millisecond magnetars with $B_{\rm p}=10^{14}\,{\rm G}$ and tidal disruptions. We also note that pion cascades could contribute to secondary energy losses. In our case studies, they play a minor role, as discussed  in appendix~\ref{app:photohadronic_secondaries}.

\subsection{Secondary acceleration}

Our case studies demonstrate that secondary acceleration can impact the neutrino spectrum: the neutrino flux at the highest energies can increase, as well as the peak energy of the neutrino spectrum. These effects vary as a function of the competition between secondary acceleration and losses, and as a function of $p_{\rm esc}$. Among our case studies, we observe mostly two different types of neutrino spectra.

When secondary acceleration is efficient ($t_{\rm acc} \ll t_{\rm loss}$), the neutrino peak energy and flux can be significantly modified. This is the case for the millisecond magnetar with $B_{\rm p}=10^{15}\,{\rm G}$ and magnetar giant flare case studies. The slope of the high-energy part of the neutrino spectrum produced through secondary acceleration is steeper than the slope of the proton spectrum, because of the impact of the energy losses of secondaries after they escape from the acceleration zone and before they decay. For instance, for magnetar giant flares, this slope is approximately $\alpha+1.5$ for $\nu_{\pi^\pm \rightarrow \nu_{\rm direct}}$ and  $\alpha+1.8$ for $\nu_{\mu^\pm \rightarrow \nu_{\rm direct}}$, where $\alpha$ is the slope of the proton spectrum. Therefore, for $p_{\rm esc,low}$, a secondary peak is produced in the neutrino spectrum, around the maximum energies predicted in section~\ref{sec:Emax}. Some neutrinos reach slightly higher energies due decay kinematics. We note that the spectra of accelerated charged pions and muons can show sharp peaks, as the spectra of accelerated protons, but these peaks are smoothed by decay kinematics, as noticed in \cite{Winter14}. For millisecond magnetars with $B_{\rm p}=10^{15}\,{\rm G}$ and magnetar giant flares, the maximum energy with secondary acceleration, and thus the energy of the secondary peak, is about $10-100$ times higher than the one predicted without secondary acceleration. For $p_{\rm esc,low}$, the secondary peak is at the same level than the primary peak, and could therefore enhance significantly the neutrino detectability at high energies. For $p_{\rm esc,high}$, the secondary peak is low but the high-energy part of the spectrum is still enhanced. The secondary peak is two orders of magnitude below the peak flux, even if it is a factor of $10$ higher than the flux obtained without secondary acceleration. We note that despite the larger decay time of charged muons and thus their greater propensity to be accelerated, the secondary peak is dominated by muon neutrinos from pion decay, due to the slope of the muon spectrum.

When secondary acceleration is less efficient, the increase of the peak energy is less pronounced and thus no secondary neutrino peak is produced. This is the case for millisecond magnetars with $B_{\rm p}=10^{14}\,{\rm G}$ and jetted tidal disruptions, for which only charged muons are expected to experience significant acceleration. Nevertheless, the neutrino flux at peak can increase by a factor of $3-5$ with respect to the case without secondary acceleration. As previously, the spectral shape is more impacted by secondary acceleration for $p_{\rm esc,low}$, as more secondaries pile-up at the highest energies. In these cases, charged pions and muons contribute equally to the neutrino peak, producing a different flavor ratio than in the case of efficient acceleration.

In our calculations of the neutrino maximum energy and spectrum, we have maximized the effects of secondary acceleration by considering the Bohm regime, together with the constant escape probability $p_{\rm esc,low}$. However, as mentioned in section~\ref{sec:model}, a large variety of models could be considered for the acceleration timescale. In the case of scattering against magnetic inhomogeneities, an analytic solution of equation~\ref{Eq:energy} is more difficult to establish. However, this differential equation can be solved numerically and several salient points can be examined despite longer computational times. As expected, the coherence length of the magnetic field $l_B$ influences the efficiency of acceleration. We note that considering a coherence length equal to the Larmor radius of the highest energy particles in the Bohm regime $r_{\rm L} = E_{\rm max}/eB$, where $E_{\rm max}=(\sqrt{\Delta}-A_1)/2 A_2$, allows to retrieve particle acceleration as efficient as in the Bohm regime. However, this coherence length that maximizes acceleration is not the same for protons, charged pions and charged muons. Therefore, if the coherence length is fixed, the efficiency of primary or secondary acceleration can be severely reduced. Nevertheless, as shown in figure~\ref{fig:Enumax}, if secondary acceleration operates in the Bohm regime the maximum energies of charged pions and muons are similar. Therefore, if the coherence length is fixed to the charged pion or charged muon maximum Larmor radius, the maximum energies of neutrinos should be similar than the ones obtained in the Bohm regime. We have confirmed this hypothesis by detailed calculations of the neutrino maximum energies, and for our case studies of the neutrino spectra. For the magnetar giant flare case study, due to the large secondary energy losses, the coherence length equal to the maximum Larmor radius of charged pions corresponds approximately to a fraction $5 \times 10^{-5}$ of the comoving size of the flaring region. We note that while increasing secondary acceleration, this coherence length decrease the proton maximum energy, and could thus impact the spectra of accelerated, interacting and escaping protons.

\section{Discussion}
\label{sec:discussion}

We have examined several physical processes that can influence the production of high-energy neutrino flares in explosive transients, with a focus on the acceleration of secondary particles, namely charged pions and muons, before they decay and produce neutrinos. Following \cite{Guepin17}, we have considered photohadronic interactions as the dominant mechanism for the production of high-energy neutrino flares. Our one-zone model can be applied to a large variety of explosive transients and enables quick parameter space scans, by varying the variability timescale $t_{\rm var}$, the bolometric luminosity $L_{\rm bol}$ and the bulk Lorentz factor $\Gamma$ characterizing the transient emission. We have calculated the maximum energy of neutrinos in this parameter space. We have also carried several case studies in order to examine in more detail the effect of secondary acceleration on high-energy neutrino spectra, and identify in which cases the maximum energy corresponds to a prominent spectral feature. For this purpose, we adopted a simple modeling of the photon spectrum, the target for photohadronic interaction: we considered a broken power-law, characterized by its break energy and two spectral indices.

We have shown that secondary acceleration have a strong impact on sources experiencing large secondary losses, and that efficient secondary acceleration can increase significantly the maximum neutrino energy, and in some cases can produce an additional spectral peak. Given the sensitivities of current neutrino observatories, the scarcity of observations, and the number of parameters involved in the modeling of transient neutrino emissions, it is still difficult to draw a definite conclusion about a specific source category. In the long term, the association of several neutrinos with one transient source and eventually the reconstruction of its neutrino spectrum, together with its gamma rays spectrum, will be required to clearly identify features of secondary acceleration. Given the peak neutrino energies predicted and the sensitivities of current high-energy neutrino detectors, these features could be probed by the next generation of neutrino detectors, such as IceCube-Gen2, or future observatories designed particularly for the detection of very-high-energy neutrinos above $10^{17}-10^{18}\,{\rm eV}$ and plan to improve significantly current sensitivity limits, as POEMMA or GRAND.

Our results are consistent with \cite{Reynoso14, Winter14} in terms of potential secondary acceleration in gamma-ray bursts, typical spectral features produced by secondary acceleration and increase of the maximum flux. However, in our work, we did not focus on a specific source class and aimed at a general model to scan the full parameter space of explosive transients. As a consequence, our model is slightly different, as we have considered a one-zone model, without distinguishing between acceleration and radiation zone. Thus we did not model explicitly particle transport. Furthermore, we have considered only one acceleration process, without distinguishing between shock and stochastic acceleration. Assuming a mono-energetic proton injection, we have treated the acceleration of protons and secondary self-consistently, with a parametrization producing power-law spectra. We accounted for the impact of photohadronic interactions on proton acceleration, which can have a strong impact on the energy required for acceleration and the radiated energy. We also evaluated the potential secondary energy losses due to pion cascades, which could be important for several source categories, such as high-luminosity gamma-ray bursts. Such estimates should be refined on a case-by-case basis, with a precise modeling of the target photon spectra and on the photopion cross section and inelasticity, in particular in the resonance region. Moreover, several proton injection spectra could be tested and additional phenomenological models for particle acceleration could be explored, for instance with a variable escape probability $p_{\rm esc}$ depending on the particle properties.

We have focused on high-energy neutrino production and did not study the associated non-thermal radiation. We note that for our case studies, the gamma-ray spectra produced through $\pi^0$ decay, and their interaction with the flaring photon background through $\gamma \gamma \rightarrow e^+ e^-$ processes, result in pile-up of gamma-ray photons in MeV or GeV energy ranges. Furthermore, we have noticed with simple estimates that secondary acceleration can increase non-thermal radiation, for instance through an increase of synchrotron radiation of charged pions and muons when they are accelerated. In most of the cases, charged muons radiate more that charged pions due to their longer lifetimes. With secondary acceleration, the total energy radiated through synchrotron can increase up to a factor $10^2$. The case studies with large secondary energy losses and efficient acceleration, namely millisecond magnetars with $B=10^{15}\,{\rm G}$ and magnetar giant flares with $p_{\rm esc,low}$, show the largest increase in radiated energy. In this cases, it could further constrain the parameter $\eta_{\rm p}$ used for spectra normalization, as it would be essential to include the contribution of secondary mesons and leptons to the total energy radiated $\sim L_{\rm bol} t_{\rm dur}$. In the future, the impact of secondary acceleration on the non-thermal radiation spectra could be tested for specific transient sources. 

\section*{Acknowledgment}
The author thank the anonymous referee for insightful feedback on this work. The author would also like to thank K. Kotera and M. Petropoulou for helpful comments and discussions. This work was supported by a fellowship from the CFM Foundation for Research, the Labex ILP (reference ANR-10-LABX-63, ANR-11-IDEX-0004-02) and the Neil Gehrels Prize Postdoctoral Fellowship.

\bibliographystyle{aa}
\bibliography{NF_reacc}

\begin{thebibliography}{50}
\expandafter\ifx\csname natexlab\endcsname\relax\def\natexlab#1{#1}\fi

\bibitem[{{Aartsen} {et~al.}(2013{\natexlab{a}}){Aartsen}, {Abbasi}, {Abdou},
  {Ackermann}, {Adams}, {Aguilar}, {Ahlers}, {Altmann}, {Auffenberg}, {Bai}, \&
  et~al.}]{Aartsen13a}
{Aartsen}, M.~G., {Abbasi}, R., {Abdou}, Y., {et~al.} ({IceCube
  Collaboration}). 2013{\natexlab{a}}, Physical Review Letters, 111, 021103

\bibitem[{{Aartsen} {et~al.}(2013{\natexlab{b}}){Aartsen}, {Abbasi}, {Abdou},
  {Ackermann}, {Adams}, {Aguilar}, {Ahlers}, {Altmann}, {Auffenberg}, {Bai}, \&
  et~al.}]{Aartsen13b}
{Aartsen}, M.~G., {Abbasi}, R., {Abdou}, Y., {et~al.} ({IceCube
  Collaboration}). 2013{\natexlab{b}}, \apj, 779, 132

\bibitem[{{Aartsen} {et~al.}(2015{\natexlab{a}}){Aartsen}, {Abraham},
  {Ackermann}, {Adams}, {Aguilar}, {Ahlers}, {Ahrens}, {Altmann}, {Anderson},
  {Ansseau}, {Anton}, {Archinger}, {Arguelles}, {Arlen}, {Auffenberg}, {Axani},
  {Bai}, {Bartos}, {Barwick}, {Baum}, {Bay}, {Beatty}, {Becker Tjus}, {Becker},
  {Beiser}, {BenZvi}, {Berghaus}, {Berley}, {Bernardini}, {Bernhard}, {Besson},
  {Binder}, {Bindig}, {Bissok}, {Blaufuss}, {Blumenthal}, {Boersma}, {Bohm},
  {B{\"o}rner}, {Bos}, {Bose}, {B{\"o}ser}, {Botner}, {Braun}, {Brayeur},
  {Bretz}, {Buzinsky}, {Casey}, {Casier}, {Cheung}, {Chirkin}, {Christov},
  {Clark}, {Classen}, {Coenders}, {Collin}, {Conrad}, {Cowen}, {Cruz Silva},
  {Daughhetee}, {Davis}, {Day}, {de Andr{\'e}}, {De Clercq}, {del Pino
  Rosendo}, {Dembinski}, {De Ridder}, {Desiati}, {de Vries}, {de Wasseige}, {de
  With}, {DeYoung}, {Diaz-V{\'e}lez}, {di Lorenzo}, {Dumm}, {Dunkman}, {Eagan},
  {Eberhardt}, {Ehrhardt}, {Eichmann}, {Euler}, {Evans}, {Evenson}, {Fadiran},
  {Fahey}, {Fazely}, {Fedynitch}, {Feintzeig}, {Felde}, {Filimonov}, {Finley},
  {Fischer-Wasels}, {Flis}, {F{\"o}sig}, {Fuchs}, {Gaisser}, {Gaior},
  {Gallagher}, {Gerhardt}, {Ghorbani}, {Gier}, {Gladstone}, {Glagla},
  {Gl{\"u}senkamp}, {Goldschmidt}, {Golup}, {Gonzalez}, {G{\'o}ra}, {Grant},
  {Groh}, {Gro{\ss}}, {Ha}, {Haack}, {Haj Ismail}, {Hallgren}, {Halzen},
  {Hansmann}, {Hanson}, {Haugen}, {Hebecker}, {Heereman}, {Helbing},
  {Hellauer}, {Hellwig}, {Hickford}, {Hignight}, {Hill}, {Hoffman}, {Hoffmann},
  {Holzapfel}, {Homeier}, {Hoshina}, {Huang}, {Huber}, {Huelsnitz}, {Hulth},
  {Hultqvist}, {In}, {Ishihara}, {Jacobi}, {Japaridze}, {Jero}, {Jones},
  {Jurkovic}, {Kalekin}, {Kaminsky}, {Kappes}, {Karg}, {Karle}, {Katori},
  {Katz}, {Kauer}, {Keivani}, {Kelley}, {Kemp}, {Kheirandish}, {Kiryluk},
  {Kl{\"a}s}, {Klein}, {Kohnen}, {Koirala}, {Kolanoski}, {Konietz}, {Koob},
  {K{\"o}pke}, {Kopper}, {Kopper}, {Koskinen}, {Kowalski}, {Krauss}, {Krings},
  {Kroll}, {Kroll}, {Kunnen}, {Kurahashi}, {Kuwabara}, {Labare}, {Lanfranchi},
  {Larson}, {Lesiak-Bzdak}, {Leuermann}, {Leuner}, {LoSecco}, {Lu},
  {L{\"u}nemann}, {Madsen}, {Maggi}, {Mahn}, {Marka}, {Marka}, {Maruyama},
  {Mase}, {Matis}, {Maunu}, {McNally}, {Meagher}, {Medici}, {Meli}, {Menne},
  {Merino}, {Meures}, {Miarecki}, {Middell}, {Middlemas}, {Mohrmann},
  {Montaruli}, {Moore}, {Morse}, {Nahnhauer}, {Naumann}, {Neer},
  {Niederhausen}, {Nowicki}, {Nygren}, {Obertacke}, {Olivas}, {Omairat},
  {O'Murchadha}, {Palazzo}, {Palczewski}, {Pand ya}, {Paul}, {Pepper},
  {P{\'e}rez de los Heros}, {Petersen}, {Pfendner}, {Pieloth}, {Pinat},
  {Pinfold}, {Posselt}, {Price}, {Przybylski}, {P{\"u}tz}, {Quinnan}, {Raab},
  {R{\"a}del}, {Rameez}, {Rawlins}, {Reimann}, {Relich}, {Resconi}, {Rhode},
  {Richman}, {Richter}, {Riedel}, {Robertson}, {Rongen}, {Rott}, {Ruhe},
  {Ryckbosch}, {Saba}, {Sabbatini}, {Sand er}, {Sandrock}, {Sandroos},
  {Sandstrom}, {Sarkar}, {Schatto}, {Scheriau}, {Schimp}, {Schmidt}, {Schmitz},
  {Schoenen}, {Sch{\"o}neberg}, {Sch{\"o}nwald}, {Schulte}, {Seckel},
  {Seunarine}, {Shaevitz}, {Shanidze}, {Smith}, {Soldin},
  {S{\"o}ldner-Rembold}, {Song}, {Spiczak}, {Spiering}, {Stahlberg},
  {Stamatikos}, {Stanev}, {Stanisha}, {Stasik}, {Stezelberger}, {Stokstad},
  {St{\"o}{\ss}l}, {Str{\"o}m}, {Strotjohann}, {Sullivan}, {Sutherland},
  {Taavola}, {Taboada}, {Taketa}, {Tanaka}, {Ter-Antonyan}, {Terliuk},
  {Te{\v{s}}i{\'c}}, {Tilav}, {Toale}, {Tobin}, {Toscano}, {Tosi},
  {Tselengidou}, {Turcati}, {Unger}, {Usner}, {Vallecorsa}, {Vandenbroucke},
  {van Eijndhoven}, {Vanheule}, {van Santen}, {Veenkamp}, {Vehring}, {Voge},
  {Vraeghe}, {Walck}, {Wallace}, {Wallraff}, {Wandkowsky}, {Weaver}, {Wendt},
  {Westerhoff}, {Whelan}, {Whitehorn}, {Wichary}, {Wiebe}, {Wiebusch}, {Wille},
  {Williams}, {Wissing}, {Wolf}, {Wood}, {Woschnagg}, {Wren}, {Xu}, {Xu}, {Xu},
  {Yanez}, {Yodh}, {Yoshida}, \& {Zoll}}]{IceCubeGen2_2015}
{Aartsen}, M.~G., {Abraham}, K., {Ackermann}, M., {et~al.} ({IceCube-Gen2
  Collaboration}). 2015{\natexlab{a}}, ArXiv e-prints
  [\eprint[arXiv]{1510.05228}]

\bibitem[{{Aartsen} {et~al.}(2018){Aartsen}, {Ackermann}, {Adams}, {Aguilar},
  {Ahlers}, {Ahrens}, {Al Samarai}, {Altmann}, {Andeen}, {Anderson}, {Ansseau},
  {Anton}, {Arg{\"u}elles}, {Auffenberg}, {Axani}, {Bagherpour}, {Bai},
  {Barron}, {Barwick}, {Baum}, {Bay}, {Beatty}, {Becker Tjus}, {Becker},
  {BenZvi}, {Berley}, {Bernardini}, {Besson}, {Binder}, {Bindig}, {Blaufuss},
  {Blot}, {Bohm}, {B{\"o}rner}, {Bos}, {B{\"o}ser}, {Botner}, {Bourbeau},
  {Bourbeau}, {Bradascio}, {Braun}, {Brenzke}, {Bretz}, {Bron},
  {Brostean-Kaiser}, {Burgman}, {Busse}, {Carver}, {Cheung}, {Chirkin},
  {Christov}, {Clark}, {Classen}, {Coenders}, {Collin}, {Conrad}, {Coppin},
  {Correa}, {Cowen}, {Cross}, {Dave}, {Day}, {de Andr{\'e}}, {De Clercq},
  {DeLaunay}, {Dembinski}, {De Ridder}, {Desiati}, {de Vries}, {de Wasseige},
  {de With}, {DeYoung}, {D{\'\i}az-V{\'e}lez}, {di Lorenzo}, {Dujmovic},
  {Dumm}, {Dunkman}, {Dvorak}, {Eberhardt}, {Ehrhardt}, {Eichmann}, {Eller},
  {Evenson}, {Fahey}, {Fazely}, {Felde}, {Filimonov}, {Finley}, {Flis},
  {Franckowiak}, {Friedman}, {Fritz}, {Gaisser}, {Gallagher}, {Gerhardt},
  {Ghorbani}, {Glauch}, {Gl{\"u}senkamp}, {Goldschmidt}, {Gonzalez}, {Grant},
  {Griffith}, {Haack}, {Hallgren}, {Halzen}, {Hanson}, {Hebecker}, {Heereman},
  {Helbing}, {Hellauer}, {Hickford}, {Hignight}, {Hill}, {Hoffman}, {Hoffmann},
  {Hoinka}, {Hokanson-Fasig}, {Hoshina}, {Huang}, {Huber}, {Hultqvist},
  {H{\"u}nnefeld}, {Hussain}, {In}, {Iovine}, {Ishihara}, {Jacobi},
  {Japaridze}, {Jeong}, {Jero}, {Jones}, {Kalaczynski}, {Kang}, {Kappes},
  {Kappesser}, {Karg}, {Karle}, {Katz}, {Kauer}, {Keivani}, {Kelley},
  {Kheirandish}, {Kim}, {Kim}, {Kintscher}, {Kiryluk}, {Kittler}, {Klein},
  {Koirala}, {Kolanoski}, {K{\"o}pke}, {Kopper}, {Kopper}, {Koschinsky},
  {Koskinen}, {Kowalski}, {Krings}, {Kroll}, {Kr{\"u}ckl}, {Kunwar},
  {Kurahashi}, {Kuwabara}, {Kyriacou}, {Labare}, {Lanfranchi}, {Larson},
  {Lauber}, {Leonard}, {Lesiak-Bzdak}, {Leuermann}, {Liu}, {Lozano Mariscal},
  {Lu}, {L{\"u}nemann}, {Luszczak}, {Madsen}, {Maggi}, {Mahn}, {Mancina},
  {Maruyama}, {Mase}, {Maunu}, {Meagher}, {Medici}, {Meier}, {Menne}, {Merino},
  {Meures}, {Miarecki}, {Micallef}, {Moment{\'e}}, {Montaruli}, {Moore}, {S},
  {Morse}, {Moulai}, {Nahnhauer}, {Nakarmi}, {Naumann}, {Neer}, {Niederhausen},
  {Nowicki}, {Nygren}, {Obertacke Pollmann}, {Olivas}, {O'Murchadha},
  {O'Sullivan}, {Palczewski}, {Pandya}, {Pankova}, {Peiffer}, {Pepper},
  {P{\'e}rez de los Heros}, {Pieloth}, {Pinat}, {Plum}, {Price}, {Przybylski},
  {Raab}, {R{\"a}del}, {Rameez}, {Rauch}, {Rawlins}, {Rea}, {Reimann},
  {Relethford}, {Relich}, {Resconi}, {Rhode}, {Richman}, {Robertson}, {Rongen},
  {Rott}, {Ruhe}, {Ryckbosch}, {Rysewyk}, {Safa}, {S{\"a}lzer}, {Sanchez
  Herrera}, {Sandrock}, {Sandroos}, {Santander}, {Sarkar}, {Sarkar},
  {Satalecka}, {Schlunder}, {Schmidt}, {Schneider}, {Schoenen},
  {Sch{\"o}neberg}, {Schumacher}, {Sclafani}, {Seckel}, {Seunarine},
  {Soedingrekso}, {Soldin}, {Song}, {Spiczak}, {Spiering}, {Stachurska},
  {Stamatikos}, {Stanev}, {Stasik}, {Stein}, {Stettner}, {Steuer},
  {Stezelberger}, {Stokstad}, {St{\"o}{\ss}l}, {Strotjohann}, {Stuttard},
  {Sullivan}, {Sutherland}, {Taboada}, {Tatar}, {Tenholt}, {Ter-Antonyan},
  {Terliuk}, {Tilav}, {Toale}, {Tobin}, {Toennis}, {Toscano}, {Tosi},
  {Tselengidou}, {Tung}, {Turcati}, {Turley}, {Ty}, {Unger}, {Usner},
  {Vandenbroucke}, {Van Driessche}, {van Eijk}, {van Eijndhoven}, {Vanheule},
  {van Santen}, {Vogel}, {Vraeghe}, {Walck}, {Wallace}, {Wallraff}, {Wandler},
  {Wandkowsky}, {Waza}, {Weaver}, {Weiss}, {Wendt}, {Werthebach}, {Westerhoff},
  {Whelan}, {Whitehorn}, {Wiebe}, {Wiebusch}, {Wille}, {Williams}, {Wills},
  {Wolf}, {Wood}, {Wood}, {Woschnagg}, {Xu}, {Xu}, {Xu}, {Yanez}, {Yodh},
  {Yoshida}, {Yuan}, {Fermi-LAT Collaboration}, {Abdollahi}, {Ajello},
  {Angioni}, {Baldini}, {Ballet}, {Barbiellini}, {Bastieri}, {Bechtol},
  {Bellazzini}, {Berenji}, {Bissaldi}, {Blandford}, {Bonino}, {Bottacini},
  {Bregeon}, {Bruel}, {Buehler}, {Burnett}, {Burns}, {Buson}, {Cameron},
  {Caputo}, {Caraveo}, {Cavazzuti}, {Charles}, {Chen}, {Cheung}, {Chiang},
  {Chiaro}, {Ciprini}, {Cohen-Tanugi}, {Conrad}, {Costantin}, {Cutini},
  {D'Ammando}, {de Palma}, {Digel}, {Di Lalla}, {Di Mauro}, {Di Venere},
  {Dom{\'\i}nguez}, {Favuzzi}, {Franckowiak}, {Fukazawa}, {Funk}, {Fusco},
  {Gargano}, {Gasparrini}, {Giglietto}, {Giomi}, {Giommi}, {Giordano},
  {Giroletti}, {Glanzman}, {Green}, {Grenier}, {Grondin}, {Guiriec}, {Harding},
  {Hayashida}, {Hays}, {Hewitt}, {Horan}, {J{\'o}hannesson}, {Kadler},
  {Kensei}, {Kocevski}, {Krauss}, {Kreter}, {Kuss}, {La Mura}, {Larsson},
  {Latronico}, {Lemoine-Goumard}, {Li}, {Longo}, {Loparco}, {Lovellette},
  {Lubrano}, {Magill}, {Maldera}, {Malyshev}, {Manfreda}, {Mazziotta},
  {McEnery}, {Meyer}, {Michelson}, {Mizuno}, {Monzani}, {Morselli},
  {Moskalenko}, {Negro}, {Nuss}, {Ojha}, {Omodei}, {Orienti}, {Orlando},
  {Palatiello}, {Paliya}, {Perkins}, {Persic}, {Pesce-Rollins}, {Piron},
  {Porter}, {Principe}, {Rain{\`o}}, {Rando}, {Rani}, {Razzano}, {Razzaque},
  {Reimer}, {Reimer}, {Renault-Tinacci}, {Ritz}, {Rochester}, {Saz Parkinson},
  {Sgr{\`o}}, {Siskind}, {Spandre}, {Spinelli}, {Suson}, {Tajima}, {Takahashi},
  {Tanaka}, {Thayer}, {Thompson}, {Tibaldo}, {Torres}, {Torresi}, {Tosti},
  {Troja}, {Valverde}, {Vianello}, {Vogel}, {Wood}, {Wood}, {Zaharijas}, {MAGIC
  Collaboration}, {Ahnen}, {Ansoldi}, {Antonelli}, {Arcaro}, {Baack},
  {Babi{\'c}}, {Banerjee}, {Bangale}, {Barres de Almeida}, {Barrio}, {Becerra
  Gonz{\'a}lez}, {Bednarek}, {Bernardini}, {Berti}, {Bhattacharyya}, {Biland},
  {Blanch}, {Bonnoli}, {Carosi}, {Carosi}, {Ceribella}, {Chatterjee}, {Colak},
  {Colin}, {Colombo}, {Contreras}, {Cortina}, {Covino}, {Cumani}, {Da Vela},
  {Dazzi}, {De Angelis}, {De Lotto}, {Delfino}, {Delgado}, {Di Pierro},
  {Dom{\'\i}nguez}, {Dominis Prester}, {Dorner}, {Doro}, {Einecke},
  {Elsaesser}, {Fallah Ramazani}, {Fern{\'a}ndez-Barral}, {Fidalgo}, {Foffano},
  {Pfrang}, {Fonseca}, {Font}, {Franceschini}, {Fruck}, {Galindo}, {Gallozzi},
  {Garc{\'\i}a L{\'o}pez}, {Garczarczyk}, {Gaug}, {Giammaria}, {Godinovi{\'c}},
  {Gora}, {Guberman}, {Hadasch}, {Hahn}, {Hassan}, {Hayashida}, {Herrera},
  {Hose}, {Hrupec}, {Inoue}, {Ishio}, {Konno}, {Kubo}, {Kushida}, {Lelas},
  {Lindfors}, {Lombardi}, {Longo}, {L{\'o}pez}, {Maggio}, {Majumdar},
  {Makariev}, {Maneva}, {Manganaro}, {Mannheim}, {Maraschi}, {Mariotti},
  {Mart{\'\i}nez}, {Masuda}, {Mazin}, {Minev}, {M}, {Mirzoyan}, {Moralejo},
  {Moreno}, {Moretti}, {Nagayoshi}, {Neustroev}, {Niedzwiecki}, {Nievas
  Rosillo}, {Nigro}, {Nilsson}, {Ninci}, {Nishijima}, {Noda}, {Nogu{\'e}s},
  {Paiano}, {Palacio}, {Paneque}, {Paoletti}, {Paredes}, {Pedaletti},
  {Peresano}, {Persic}, {Prada Moroni}, {Prandini}, {Puljak}, {Rodriguez
  Garcia}, {Reichardt}, {Rhode}, {Rib{\'o}}, {Rico}, {Righi}, {Rugliancich},
  {Saito}, {Satalecka}, {Schweizer}, {Sitarek}, {{\v{S}}nidaric ́},
  {Sobczynska}, {Stamerra}, {Strzys}, {Suri{\'c}}, {Takahashi}, {Tavecchio},
  {Temnikov}, {Terzi{\'c}}, {Teshima}, {Torres-Alb{\`a}}, {Treves},
  {Tsujimoto}, {Vanzo}, {Vazquez Acosta}, {Vovk}, {Ward}, {Will}, {S}, {Zaric
  ́}, {AGILE Team}, {Lucarelli}, {Tavani}, {Piano}, {Donnarumma}, {Pittori},
  {Verrecchia}, {Barbiellini}, {Bulgarelli}, {Caraveo}, {Cattaneo},
  {Colafrancesco}, {Costa}, {Di Cocco}, {Ferrari}, {Gianotti}, {Giuliani},
  {Lipari}, {Mereghetti}, {Morselli}, {Pacciani}, {Paoletti}, {Parmiggiani},
  {Pellizzoni}, {Picozza}, {Pilia}, {Rappoldi}, {Trois}, {Vercellone},
  {Vittorini}, {ASAS-SN Team}, {Stanek}, {Kochanek}, {Beacom}, {Thompson},
  {Holoien}, {Dong}, {Prieto}, {Shappee}, {Holmbo}, {HAWC Collaboration},
  {Abeysekara}, {Albert}, {Alfaro}, {Alvarez}, {Arceo},
  {Arteaga-Vel{\'a}zquez}, {Avila Rojas}, {Ayala Solares}, {Becerril},
  {Belmont-Moreno}, {Bernal}, {Caballero-Mora}, {Capistr{\'a}n},
  {Carrami{\~n}ana}, {Casanova}, {Castillo}, {Cotti}, {Cotzomi}, {Couti{\~n}o
  de Le{\'o}n}, {De Le{\'o}n}, {De la Fuente}, {Diaz Hernandez}, {Dichiara},
  {Dingus}, {DuVernois}, {D{\'\i}az-V{\'e}lez}, {Ellsworth}, {Engel},
  {Fiorino}, {Fleischhack}, {Fraija}, {Garc{\'\i}a-Gonz{\'a}lez}, {Garfias},
  {Gonz{\'a}lez Mu{\~n}oz}, {Gonz{\'a}lez}, {Goodman}, {Hampel-Arias},
  {Harding}, {Hernand ez}, {Hona}, {Hueyotl-Zahuantitla}, {Hui},
  {H{\"u}ntemeyer}, {Iriarte}, {Jardin-Blicq}, {Joshi}, {Kaufmann}, {Kunde},
  {Lara}, {Lauer}, {Lee}, {Lennarz}, {Le{\'o}n Vargas}, {Linnemann},
  {Longinotti}, {Luis-Raya}, {Luna-Garc{\'\i}a}, {Malone}, {Marinelli},
  {Martinez}, {Martinez-Castellanos}, {Mart{\'\i}nez-Castro},
  {Mart{\'\i}nez-Huerta}, {Matthews}, {Miranda-Romagnoli}, {Moreno},
  {Mostaf{\'a}}, {Nayerhoda}, {Nellen}, {Newbold}, {Nisa}, {Noriega-Papaqui},
  {Pelayo}, {Pretz}, {P{\'e}rez-P{\'e}rez}, {Ren}, {Rho}, {Rivi{\`e}re},
  {Rosa-Gonz{\'a}lez}, {Rosenberg}, {Ruiz-Velasco}, {Ruiz-Velasco}, {Salesa
  Greus}, {Sandoval}, {Schneider}, {Schoorlemmer}, {Sinnis}, {Smith},
  {Springer}, {Surajbali}, {Tibolla}, {Tollefson}, {Torres}, {Villase{\~n}or},
  {Weisgarber}, {Werner}, {Yapici}, {Gaurang}, {Zepeda}, {Zhou}, {{\'A}lvarez},
  {H.~E.~S.~S. Collaboration}, {Abdalla}, {Ang{\"u}ner}, {Armand}, {Backes},
  {Becherini}, {Berge}, {B{\"o}ttcher}, {Boisson}, {Bolmont}, {Bonnefoy},
  {Bordas}, {Brun}, {B{\"u}chele}, {Bulik}, {Caroff}, {Carosi}, {Casanova},
  {Cerruti}, {Chakraborty}, {Chandra}, {Chen}, {Colafrancesco}, {Davids},
  {Deil}, {Devin}, {Djannati-Ata{\"\i}}, {Egberts}, {Emery}, {Eschbach},
  {Fiasson}, {Fontaine}, {Funk}, {F{\"u}{\ss}ling}, {Gallant}, {Gat{\'e}},
  {Giavitto}, {Glawion}, {Glicenstein}, {Gottschall}, {Grondin}, {Haupt},
  {Henri}, {Hinton}, {Hoischen}, {Holch}, {Huber}, {Jamrozy}, {Jankowsky},
  {Jankowsky}, {Jouvin}, {Jung-Richardt}, {Kerszberg}, {Kh{\'e}lifi}, {King},
  {Klepser}, {Kluz ́niak}, {Komin}, {Kraus}, {Lefaucheur}, {Lemi{\`e}re},
  {Lemoine-Goumard}, {Lenain}, {Leser}, {Lohse}, {L{\'o}pez-Coto}, {Lorentz},
  {Lypova}, {Marandon}, {Guillem Mart{\'\i}-Devesa}, {Maurin}, {Mitchell},
  {Moderski}, {Mohamed}, {Mohrmann}, {Moulin}, {Murach}, {de Naurois},
  {Niederwanger}, {Niemiec}, {Oakes}, {O'Brien}, {Ohm}, {Ostrowski}, {Oya},
  {Panter}, {Parsons}, {Perennes}, {Piel}, {Pita}, {Poireau}, {Priyana Noel},
  {Prokoph}, {P{\"u}hlhofer}, {Quirrenbach}, {Raab}, {Rauth}, {Renaud},
  {Rieger}, {Rinchiuso}, {Romoli}, {Rowell}, {Rudak}, {Sasaki}, {Sanchez},
  {Schlickeiser}, {Sch{\"u}ssler}, {Schulz}, {Schwanke}, {Seglar-Arroyo},
  {Shafi}, {Simoni}, {Sol}, {Stegmann}, {Steppa}, {Tavernier}, {Taylor},
  {Tiziani}, {Trichard}, {Tsirou}, {van Eldik}, {van Rensburg}, {van Soelen},
  {Veh}, {Vincent}, {Voisin}, {Wagner}, {Wagner}, {Wierzcholska}, {Zanin},
  {Zdziarski}, {Zech}, {Ziegler}, {Zorn}, {{\.Z}ywucka}, {INTEGRAL Team},
  {Savchenko}, {Ferrigno}, {Bazzano}, {Diehl}, {Kuulkers}, {Laurent},
  {Mereghetti}, {Natalucci}, {Panessa}, {Rodi}, {Ubertini}, {Kanata}, Teams,
  {Morokuma}, {Ohta}, {Tanaka}, {Mori}, {Yamanaka}, {Kawabata}, {Utsumi},
  {Nakaoka}, {Kawabata}, {Nagashima}, {Yoshida}, {Matsuoka}, {Itoh}, {Kapteyn
  Team}, {Keel}, {Liverpool Telescope Team}, {Copperwheat}, {Steele},
  {Swift/NuSTAR Team}, {Cenko}, {Cowen}, {DeLaunay}, {Evans}, {Fox}, {Keivani},
  {Kennea}, {Marshall}, {Osborne}, {Santander}, {Tohuvavohu}, {Turley},
  {VERITAS Collaboration}, {Abeysekara}, {Archer}, {Benbow}, {Bird}, {Brill},
  {Brose}, {Buchovecky}, {Buckley}, {Bugaev}, {Christiansen}, {Connolly},
  {Cui}, {Daniel}, {Errando}, {Falcone}, {Feng}, {Finley}, {Fortson},
  {Furniss}, {Gueta}, {H{\"u}tten}, {Hervet}, {Hughes}, {Humensky}, {Johnson},
  {Kaaret}, {Kar}, {Kelley-Hoskins}, {Kertzman}, {Kieda}, {Krause},
  {Krennrich}, {Kumar}, {Lang}, {Lin}, {Maier}, {McArthur}, {Moriarty},
  {Mukherjee}, {Nieto}, {O'Brien}, {Ong}, {Otte}, {Park}, {Petrashyk}, {Pohl},
  {Popkow}, {Pueschel}, {Quinn}, {Ragan}, {Reynolds}, {Richards}, {Roache},
  {Rulten}, {Sadeh}, {Santander}, {Scott}, {Sembroski}, {Shahinyan}, {Sushch},
  {Tr{\'e}panier}, {Tyler}, {Vassiliev}, {Wakely}, {Weinstein}, {Wells},
  {Wilcox}, {Wilhelm}, {Williams}, {Zitzer}, {VLA/B Team}, {Tetarenko},
  {Kimball}, {Miller-Jones}, \& {Sivakoff}}]{IceCube18}
{Aartsen}, M.~G., {Ackermann}, M., {Adams}, J., {et~al.} ({Fermi-LAT}, {HAWC},
  {H.~E.~S.~S.}, {IceCube}, {MAGIC}, {VERITAS Collaborations}, {AGILE},
  {ASAS-SN}, {INTEGRAL}, {Kapteyn}, {Liverpool Telescope}, {Swift/NuSTAR}, \&
  {VLA/B Teams}). 2018, Science, 361, 1378

\bibitem[{{Aartsen} {et~al.}(2015{\natexlab{b}}){Aartsen}, {Ackermann},
  {Adams}, {Aguilar}, {Ahlers}, {Ahrens}, {Altmann}, {Anderson}, {Archinger},
  {Arguelles}, \& et~al.}]{Aartsen15}
{Aartsen}, M.~G., {Ackermann}, M., {Adams}, J., {et~al.} ({IceCube
  Collaboration}). 2015{\natexlab{b}}, \apj, 807, 46

\bibitem[{{Abbasi} {et~al.}(2012){Abbasi}, {Abdou}, {Abu-Zayyad}, {Adams},
  {Aguilar}, {Ahlers}, {Andeen}, {Auffenberg}, {Bai}, {Baker}, \&
  et~al.}]{Abbasi12a}
{Abbasi}, R., {Abdou}, Y., {Abu-Zayyad}, T., {et~al.} ({IceCube
  Collaboration}). 2012, \apj, 744, 1

\bibitem[{{Abbott} {et~al.}(2016{\natexlab{a}}){Abbott}, {Abbott}, {Abbott},
  {Abernathy}, {Acernese}, {Ackley}, {Adams}, {Adams}, {Addesso}, {Adhikari},
  \& et~al.}]{LIGO2016b}
{Abbott}, B.~P., {Abbott}, R., {Abbott}, T.~D., {et~al.} ({LIGO Scientific
  Collaboration} \& {Virgo Collaboration}). 2016{\natexlab{a}}, Physical Review
  Letters, 116, 241103

\bibitem[{{Abbott} {et~al.}(2016{\natexlab{b}}){Abbott}, {Abbott}, {Abbott},
  {Abernathy}, {Acernese}, {Ackley}, {Adams}, {Adams}, {Addesso}, {Adhikari},
  \& et~al.}]{LIGO2016a}
{Abbott}, B.~P., {Abbott}, R., {Abbott}, T.~D., {et~al.} ({LIGO Scientific
  Collaboration} \& {Virgo Collaboration}). 2016{\natexlab{b}}, Physical Review
  Letters, 116, 061102

\bibitem[{{Abbott} {et~al.}(2017){Abbott}, {Abbott}, {Abbott}, {Acernese},
  {Ackley}, {Adams}, {Adams}, {Addesso}, {Adhikari}, {Adya}, \&
  et~al.}]{LigoVirgoNSNS17}
{Abbott}, B.~P., {Abbott}, R., {Abbott}, T.~D., {et~al.} ({LIGO Scientific
  Collaboration} \& {Virgo Collaboration}). 2017, Physical Review Letters, 119,
  161101

\bibitem[{{Aguilar} {et~al.}(2019){Aguilar}, {Allison}, {Archambault},
  {Beatty}, {Besson}, {Botner}, {Buitink}, {Chen}, {Clark}, {Connolly},
  {Deaconu}, {de Kockere}, {DuVernois}, {van Eijndhoven}, {Finley}, {Garcia},
  {Hallgren}, {Halzen}, {Hanson}, {Hanson}, {P{\'e}rez de los Heros},
  {Hoffman}, {Hokanson-Fasig}, {Hughes}, {Hultqvist}, {Ishihara}, {Karle},
  {Kelley}, {Klein}, {Kowalski}, {Kravchenko}, {Latif}, {Liu}, {Lu}, {Mase},
  {Morse}, {Nam}, {Nelles}, {Oberla}, {Pfendner}, {Pan}, {Plaisier}, {Prohira},
  {Robertson}, {Rolla}, {Ryckbosch}, {Schr{\"o}der}, {Seckel}, {Shultz},
  {Smith}, {Southall}, {O'Sullivan}, {Toscano}, {Torres-Espinosa}, {Unger},
  {Vieregg}, {de Vries}, {Wang}, {Welling}, {Wissel}, \& {Yoshida}}]{RNO19}
{Aguilar}, J.~A., {Allison}, P., {Archambault}, S., {et~al.} 2019, {ArXiv
  e-prints} [\eprint[arXiv]{1907.12526}]

\bibitem[{{Alvarez-Muniz} {et~al.}(2018){Alvarez-Muniz}, {Alves Batista},
  {Balagopal V.}, {Bolmont}, {Bustamante}, {Carvalho}, {Charrier}, {Cognard},
  {Decoene}, {Denton}, {De Jong}, {De Vries}, {Engel}, {Fang}, {Finley},
  {Gabici}, {Gou}, {Gu}, {Gu{\'e}pin}, {Hu}, {Huang}, {Kotera}, {Le Coz},
  {Lenain}, {Lu}, {Martineau-Huynh}, {Mostaf{\'a}}, {Mottez}, {Murase},
  {Niess}, {Oikonomou}, {Pierog}, {Qian}, {Qin}, {Ran}, {Renault-Tinacci},
  {Roth}, {Schr{\"o}der}, {Sch{\"u}ssler}, {Tasse}, {Timmermans}, {Tueros},
  {Wu}, {Zarka}, {Zech}, {Zhang}, {Zhang}, {Zhang}, {Zheng}, \&
  {Zilles}}]{GRAND18}
{Alvarez-Muniz}, J., {Alves Batista}, R., {Balagopal V.}, A., {et~al.} ({GRAND
  Collaboration}). 2018, {ArXiv e-prints} [\eprint[arXiv]{1810.09994}]

\bibitem[{{Arons}(2003)}]{Arons03}
{Arons}, J. 2003, \apj, 589, 871

\bibitem[{{Atoyan} \& {Dermer}(2001)}]{Atoyan01}
{Atoyan}, A. \& {Dermer}, C.~D. 2001, \prl, 87, 221102

\bibitem[{{Bednarek} \& {Protheroe}(1999)}]{Bednarek99}
{Bednarek}, W. \& {Protheroe}, R.~J. 1999, \mnras, 302, 373

\bibitem[{{Biehl} {et~al.}(2018){Biehl}, {Boncioli}, {Lunardini}, \&
  {Winter}}]{Biehl17}
{Biehl}, D., {Boncioli}, D., {Lunardini}, C., \& {Winter}, W. 2018, Scientific
  Reports, 8, 10828

\bibitem[{{Blasi} {et~al.}(2000){Blasi}, {Epstein}, \& {Olinto}}]{Blasi00}
{Blasi}, P., {Epstein}, R.~I., \& {Olinto}, A.~V. 2000, \apjl, 533, L123

\bibitem[{{Bustamante} \& {Ahlers}(2019)}]{Bustamante19}
{Bustamante}, M. \& {Ahlers}, M. 2019, \prl, 122, 241101

\bibitem[{{Dai} \& {Fang}(2017)}]{Dai16}
{Dai}, L. \& {Fang}, K. 2017, \mnras, 469, 1354

\bibitem[{{Dermer} {et~al.}(2014){Dermer}, {Murase}, \& {Inoue}}]{Dermer14}
{Dermer}, C.~D., {Murase}, K., \& {Inoue}, Y. 2014, Journal of High Energy
  Astrophysics, 3, 29

\bibitem[{{Fang} {et~al.}(2012){Fang}, {Kotera}, \& {Olinto}}]{Fang12}
{Fang}, K., {Kotera}, K., \& {Olinto}, A.~V. 2012, \apj, 750, 118

\bibitem[{{Fang} {et~al.}(2013){Fang}, {Kotera}, \& {Olinto}}]{Fang13}
{Fang}, K., {Kotera}, K., \& {Olinto}, A.~V. 2013, J. Cos. and Astro. Phys., 3,
  10

\bibitem[{{Fang} \& {Metzger}(2017)}]{Fang17}
{Fang}, K. \& {Metzger}, B.~D. 2017, \apj, 849, 153

\bibitem[{{Gao} {et~al.}(2019){Gao}, {Fedynitch}, {Winter}, \& {Pohl}}]{Gao19}
{Gao}, S., {Fedynitch}, A., {Winter}, W., \& {Pohl}, M. 2019, Nature Astronomy,
  3, 88

\bibitem[{{Gu{\'e}pin} \& {Kotera}(2017)}]{Guepin17}
{Gu{\'e}pin}, C. \& {Kotera}, K. 2017, \aap, 603, A76

\bibitem[{{Gu{\'e}pin} {et~al.}(2018){Gu{\'e}pin}, {Kotera}, {Barausse},
  {Fang}, \& {Murase}}]{Guepin18b}
{Gu{\'e}pin}, C., {Kotera}, K., {Barausse}, E., {Fang}, K., \& {Murase}, K.
  2018, \aap, 616, A179

\bibitem[{{Halzen} \& {Hooper}(2005)}]{Halzen05}
{Halzen}, F. \& {Hooper}, D. 2005, Astroparticle Physics, 23, 537

\bibitem[{{H{\"u}mmer} {et~al.}(2010){H{\"u}mmer}, {R{\"u}ger}, {Spanier}, \&
  {Winter}}]{Hummer10}
{H{\"u}mmer}, S., {R{\"u}ger}, M., {Spanier}, F., \& {Winter}, W. 2010, \apj,
  721, 630

\bibitem[{{Klein} {et~al.}(2013){Klein}, {Mikkelsen}, \& {Becker
  Tjus}}]{Klein13}
{Klein}, S.~R., {Mikkelsen}, R.~E., \& {Becker Tjus}, J. 2013, \apj, 779, 106

\bibitem[{{Kotera} {et~al.}(2009){Kotera}, {Allard}, {Murase}, {Aoi}, {Dubois},
  {Pierog}, \& {Nagataki}}]{Kotera09}
{Kotera}, K., {Allard}, D., {Murase}, K., {et~al.} 2009, ApJ, 707, 370

\bibitem[{{Kotera} {et~al.}(2015){Kotera}, {Amato}, \& {Blasi}}]{Kotera15}
{Kotera}, K., {Amato}, E., \& {Blasi}, P. 2015, JCAP, 8, 026

\bibitem[{{Lemoine} {et~al.}(2015){Lemoine}, {Kotera}, \&
  {P{\'e}tri}}]{Lemoine15}
{Lemoine}, M., {Kotera}, K., \& {P{\'e}tri}, J. 2015, \jcap, 7, 016

\bibitem[{{Lemoine} \& {Waxman}(2009)}]{Lemoine09}
{Lemoine}, M. \& {Waxman}, E. 2009, Journal of Cosmology and Astro-Particle
  Physics, 2009, 009

\bibitem[{{Lunardini} \& {Winter}(2016)}]{Lunardini16}
{Lunardini}, C. \& {Winter}, W. 2016, ArXiv e-prints
  [\eprint[arXiv]{1612.03160}]

\bibitem[{{Merten} {et~al.}(2017){Merten}, {Becker Tjus}, {Fichtner},
  {Eichmann}, \& {Sigl}}]{CRPropa_v3.1}
{Merten}, L., {Becker Tjus}, J., {Fichtner}, H., {Eichmann}, B., \& {Sigl}, G.
  2017, \jcap, 2017, 046

\bibitem[{{M{\'e}sz{\'a}ros}(2015)}]{Meszaros15}
{M{\'e}sz{\'a}ros}, P. 2015, ArXiv e-prints [\eprint[arXiv]{1511.01396}]

\bibitem[{{Murase} {et~al.}(2012){Murase}, {Asano}, {Terasawa}, \&
  {M{\'e}sz{\'a}ros}}]{Murase12}
{Murase}, K., {Asano}, K., {Terasawa}, T., \& {M{\'e}sz{\'a}ros}, P. 2012,
  \apj, 746, 164

\bibitem[{{Murase} {et~al.}(2008){Murase}, {Ioka}, {Nagataki}, \&
  {Nakamura}}]{Murase08}
{Murase}, K., {Ioka}, K., {Nagataki}, S., \& {Nakamura}, T. 2008, Phys. Rev. D,
  78, 023005

\bibitem[{{Murase} \& {Nagataki}(2006)}]{Murase06_flares}
{Murase}, K. \& {Nagataki}, S. 2006, Physical Review Letters, 97, 051101

\bibitem[{{Murase} {et~al.}(2018){Murase}, {Oikonomou}, \&
  {Petropoulou}}]{Murase18}
{Murase}, K., {Oikonomou}, F., \& {Petropoulou}, M. 2018, \apj, 865, 124

\bibitem[{{Olinto} {et~al.}(2017){Olinto}, {Adams}, {Aloisio}, {Anchordoqui},
  {Bergman}, {Bertaina}, {Bertone}, {Christl}, {Csorna}, {Eser}, {Fenu},
  {Hays}, {Hunter}, {Judd}, {Jun}, {Krizmanic}, {Kuznetsov}, {Martinez-Sierra},
  {Mastafa}, {Matthews}, {McEnery}, {Mitchell}, {Neronov}, {Otte}, {Parizot},
  {Paul}, {Perkins}, {Pr{\'e}v{\^o}t}, {Reardon}, {Reno}, {Sarazin},
  {Shinozaki}, {Stecker}, {Streitmatter}, {Wiencke}, \& {Young}}]{POEMMA17}
{Olinto}, A.~V., {Adams}, J.~H., {Aloisio}, R., {et~al.} ({POEMMA
  collaboration}). 2017, ArXiv e-prints [\eprint[arXiv]{1708.07599}]

\bibitem[{{Otte} {et~al.}(2019){Otte}, {Brown}, {Doro}, {Falcone}, {Holder},
  {Judd}, {Kaaret}, {Mariotti}, {Murase}, \& {Taboada}}]{Trinity19}
{Otte}, N., {Brown}, A.~M., {Doro}, M., {et~al.} 2019, in \baas, Vol.~51, 67

\bibitem[{{Petropoulou} {et~al.}(2016){Petropoulou}, {Coenders}, \&
  {Dimitrakoudis}}]{Petropoulou16}
{Petropoulou}, M., {Coenders}, S., \& {Dimitrakoudis}, S. 2016, Astroparticle
  Physics, 80, 115

\bibitem[{{Reynoso}(2014)}]{Reynoso14}
{Reynoso}, M.~M. 2014, \aap, 564, A74

\bibitem[{{Senno} {et~al.}(2016){Senno}, {Murase}, \& {Meszaros}}]{Senno16b}
{Senno}, N., {Murase}, K., \& {Meszaros}, P. 2016, {High-energy neutrino
  flashes from x-ray bright and dark tidal disruptions events}

\bibitem[{{Venkatesan} {et~al.}(1997){Venkatesan}, {Coleman Miller}, \&
  {Olinto}}]{Venkatesan97}
{Venkatesan}, A., {Coleman Miller}, M., \& {Olinto}, A.~V. 1997, \apj, 484, 323

\bibitem[{{Wang} \& {Liu}(2016)}]{Wang16}
{Wang}, X.-Y. \& {Liu}, R.-Y. 2016, \prd, 93, 083005

\bibitem[{{Wang} {et~al.}(2011){Wang}, {Liu}, {Dai}, \& {Cheng}}]{Wang11}
{Wang}, X.-Y., {Liu}, R.-Y., {Dai}, Z.-G., \& {Cheng}, K.~S. 2011, \prd, 84,
  081301

\bibitem[{{Waxman} \& {Bahcall}(1997)}]{Waxman97_GRB}
{Waxman}, E. \& {Bahcall}, J. 1997, Physical Review Letters, 78, 2292

\bibitem[{{Winter} {et~al.}(2014){Winter}, {Becker Tjus}, \&
  {Klein}}]{Winter14}
{Winter}, W., {Becker Tjus}, J., \& {Klein}, S.~R. 2014, A\&A, 569, A58

\bibitem[{{Zhang} {et~al.}(2017){Zhang}, {Murase}, {Oikonomou}, \&
  {Li}}]{Zhang17}
{Zhang}, B.~T., {Murase}, K., {Oikonomou}, F., \& {Li}, Z. 2017, \prd, 96,
  063007

\end{thebibliography}

\appendix

\section{Pion cascades}\label{app:photohadronic_secondaries}

A large fraction of the parameter space considered can lead to efficient photomeson production. It is therefore important to assess the impact of interactions between photons and charged pions, as possible additional secondary energy losses. For this purpose, we calculate in the parameter space of explosive transients the probability that energy losses due to photopion interactions happen before synchrotron or adiabatic energy losses or decay, $p_\pi \sim \exp[-t_{\gamma \pi}(1/t_{\rm dyn} + 1/t_{\rm syn} +1/t_{\rm dec})]$, where $t_{\rm dyn}$ is the dynamical timescale, and $t_{\rm syn}$ and $t_{\rm dec}$ are respectively the synchrotron energy-loss timescale and the decay time for charged pions. To calculate the photopion energy-loss timescale $t_{\gamma \pi}$, we approximate the $\gamma\pi^{\pm}$ inelastic cross section by $2/3$ of the photomeson production cross section, and consider the same inelasticity. This approximation is more accurate in the multi-pion production region; however, for the purpose of our simple estimates, we extend this approximation until the pion production threshold and do not model precisely the $\rho(770)$ resonance region. The timescales are evaluated at the maximum pion energy in the comoving frame obtained in section~\ref{sec:Emax} without secondary acceleration.

\begin{figure}[!tp]
\vspace{1cm}
\centering
{\includegraphics[width=0.49\textwidth]{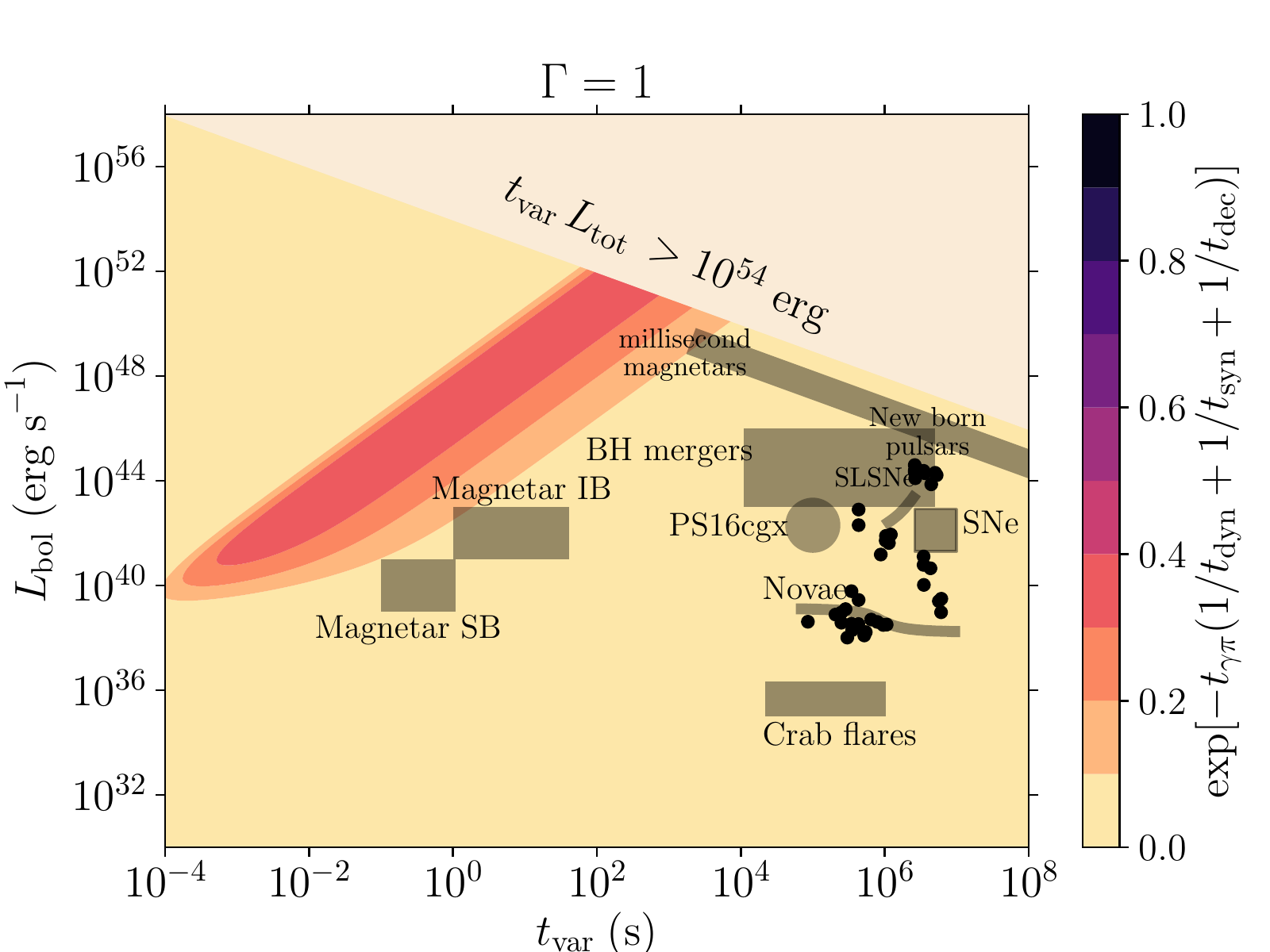}}
{\includegraphics[width=0.49\textwidth]{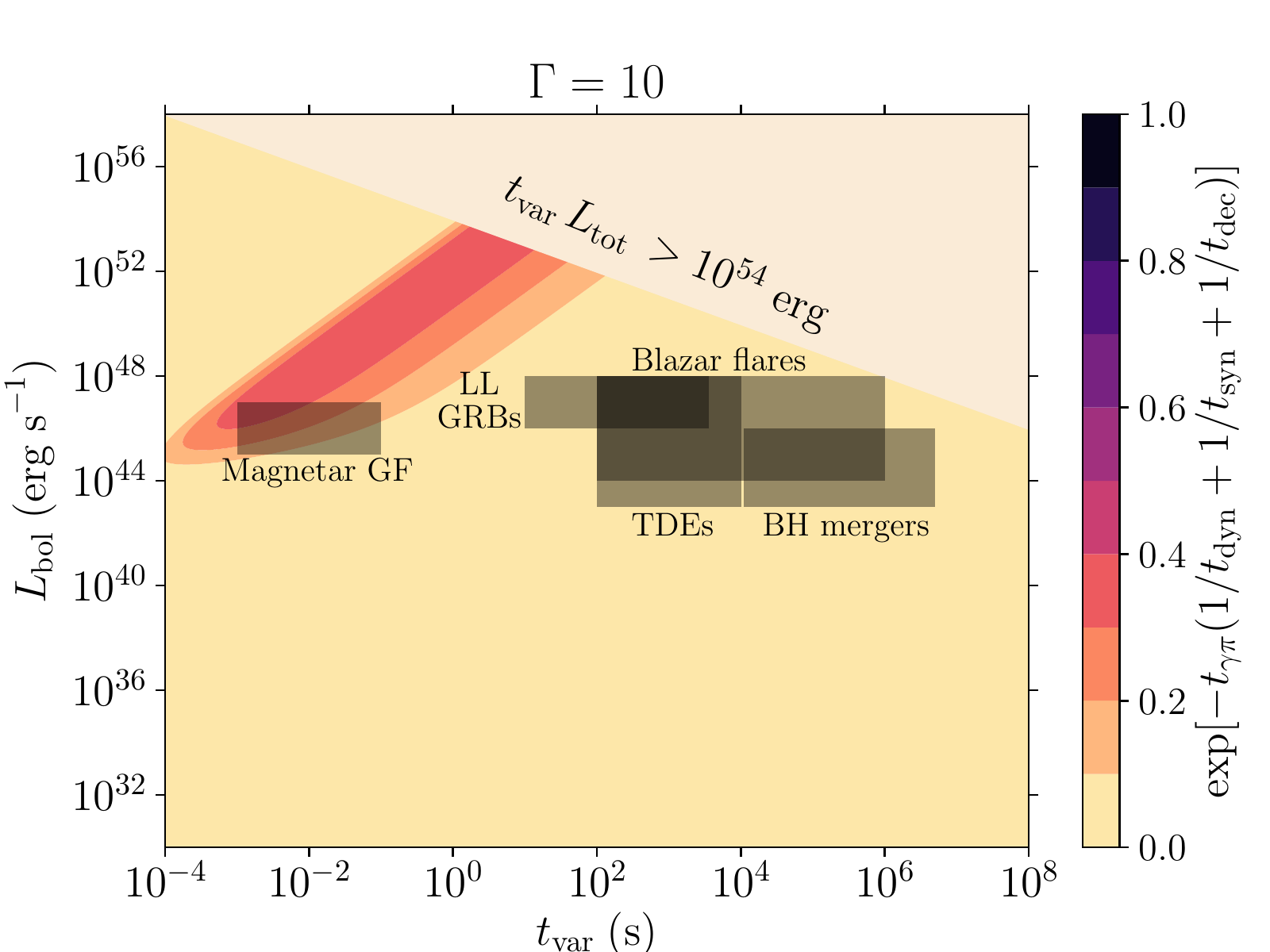}}
{\includegraphics[width=0.49\textwidth]{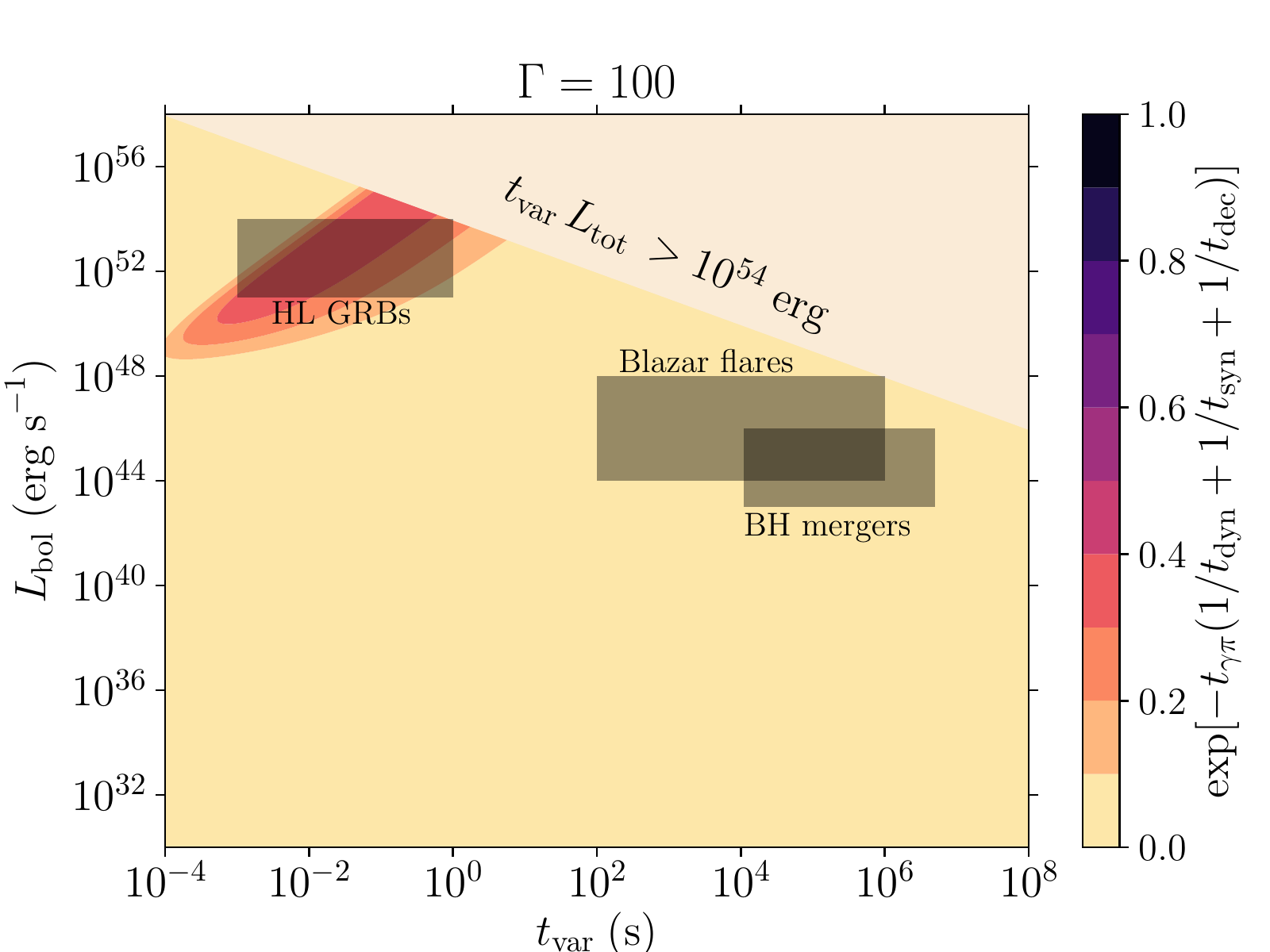}}
\caption{Probability that energy losses due to photopion interactions happen before synchrotron or adiabatic energy losses or decay, in the parameter space $t_{\rm var}-L_{\rm bol}$, for a power-law spectrum with $\epsilon_{\rm b}=10^5\,{\rm eV}$, $a=1.8$ and $b=3.1$, for $\Gamma=1,10,100$ (from top to bottom).}\label{fig:p_gpi}
\end{figure}

The results are illustrated in figure~\ref{fig:p_gpi} in the parameter space $t_{\rm var}-L_{\rm bol}$, for $\Gamma=1,10,100$. We consider a power-law spectrum with $\epsilon_{\rm b}=10^5\,{\rm eV}$, $a=1.8$ and $b=3.1$. For these parameters, we note that photopion interactions can have an impact in a limited fraction of the parameter space, and that this impact is small. The impacted region of the parameter space shifts toward higher variability timescales when $\epsilon_{\rm b}$ decreases. Its width towards low luminosities is controlled by $a$, and decreases when $a$ decreases. Its width towards high luminosities is controlled by $b$, and increases when $b$ increases. As expected, the photon spectrum influences strongly the impact of photopion interactions.

In the regions of the parameter space where pions cascades have an impact, charged pions are also subject to strong synchrotron losses. We recall that in this study, the parameter $\eta_B = 1$ controls the magnetic field, such that $U_B = \eta_B U_{\rm rad}$. A smaller $\eta_B$ value would decrease synchrotron losses, and pions cascades would play a more prominent role in the energy losses of charged pion. Our parameter $\eta_B = 1$ is thus maximizing synchrotron losses, and smaller $\eta_B$ would require a more careful treatment of pion cascades for the sources categories considered in section~\ref{sec:Spec}.

In figure~\ref{fig:p_gpi}, we have focused on the maximum pion energy obtained without secondary acceleration, as the $\nu_{\pi^\pm \rightarrow \nu_{\rm direct}}$ spectrum peaks usually at this energy. However, we note that a large amount of charged pions can be accelerated at larger energies and contribute to a secondary peak in the neutrino spectrum. At higher energies, photopion interactions affect a larger fraction of the parameter space, while remaining subdominant. Therefore, in the cases where secondary acceleration is efficient and produces a secondary peak in the neutrino spectrum, this secondary peak could be more affected by pion cascades than the primary peak.

To conclude, explosive transients with small variability timescales and high luminosities -such as high-luminosity gamma-ray bursts- are the most likely to be affected by pion cascades. Due to the numerous parameters affecting these interactions, case-by-case estimates are required. Among our case studies, the impact of pion cascades is small, due to the hard photon spectra in the case of magnetars, or due to the variability timescales in the case of tidal disruptions.

\section{Characteristic length scales}\label{app:MFP}

The characteristic length scales of energy loss and gain processes for protons are illustrated in figure~\ref{fig:MFP}, for the case studies detailed in section~\ref{sec:Spec}, namely millisecond magnetars with $B_{\rm p}=10^{14}\,{\rm G}$ and with $B_{\rm p}=10^{15}\,{\rm G}$, magnetar giant flares, and tidal disruptions. We show the photopion scattering length scale, the synchrotron, Bethe-Heitler and inverse Compton energy-loss length scales, the acceleration length scale and the typical comoving size of the flaring region. 

Acceleration, photopion production and synchrotron losses are the dominant processes. Among the energy loss processes, photopion production and synchrotron are dominant at the highest energies. We note that Bethe-Heitler losses can be marginally dominant at low energies, and at the lowest energies adiabatic losses (or escape from the flaring region) dominate. Inverse Compton losses are always subdominant. Due to the equipartition hypothesis between the magnetic and radiation energy densities and $\eta_B=1$, the synchrotron and inverse Compton energy-loss timescales are equal in the Thomson regime. Inverse Compton losses are inefficient at the highest energies due to the transition from the Thomson to the Klein-Nishina regime. 

\noindent
\begin{minipage}{\textwidth}%
{\includegraphics[width=0.49\textwidth]{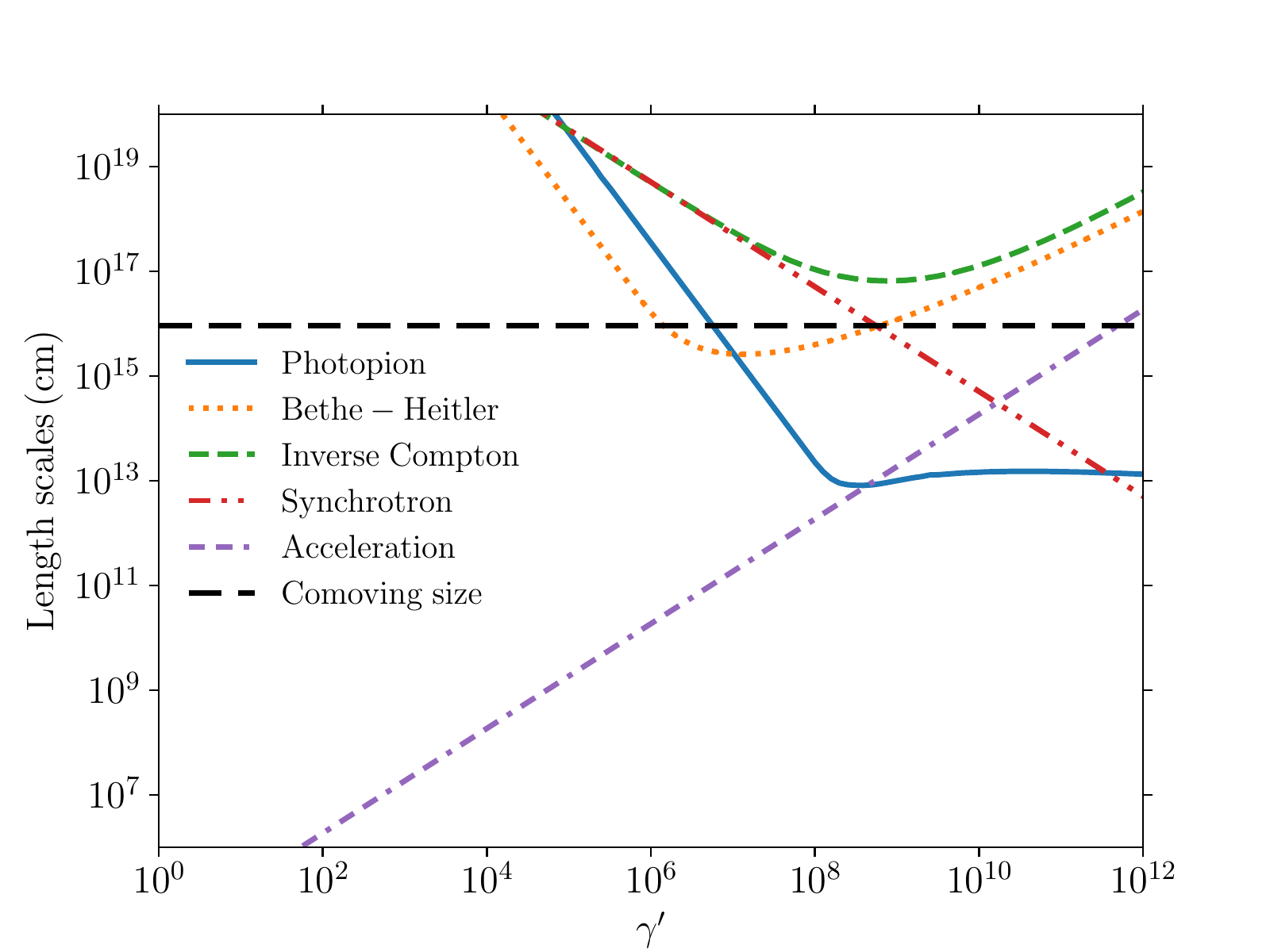}}
{\includegraphics[width=0.49\textwidth]{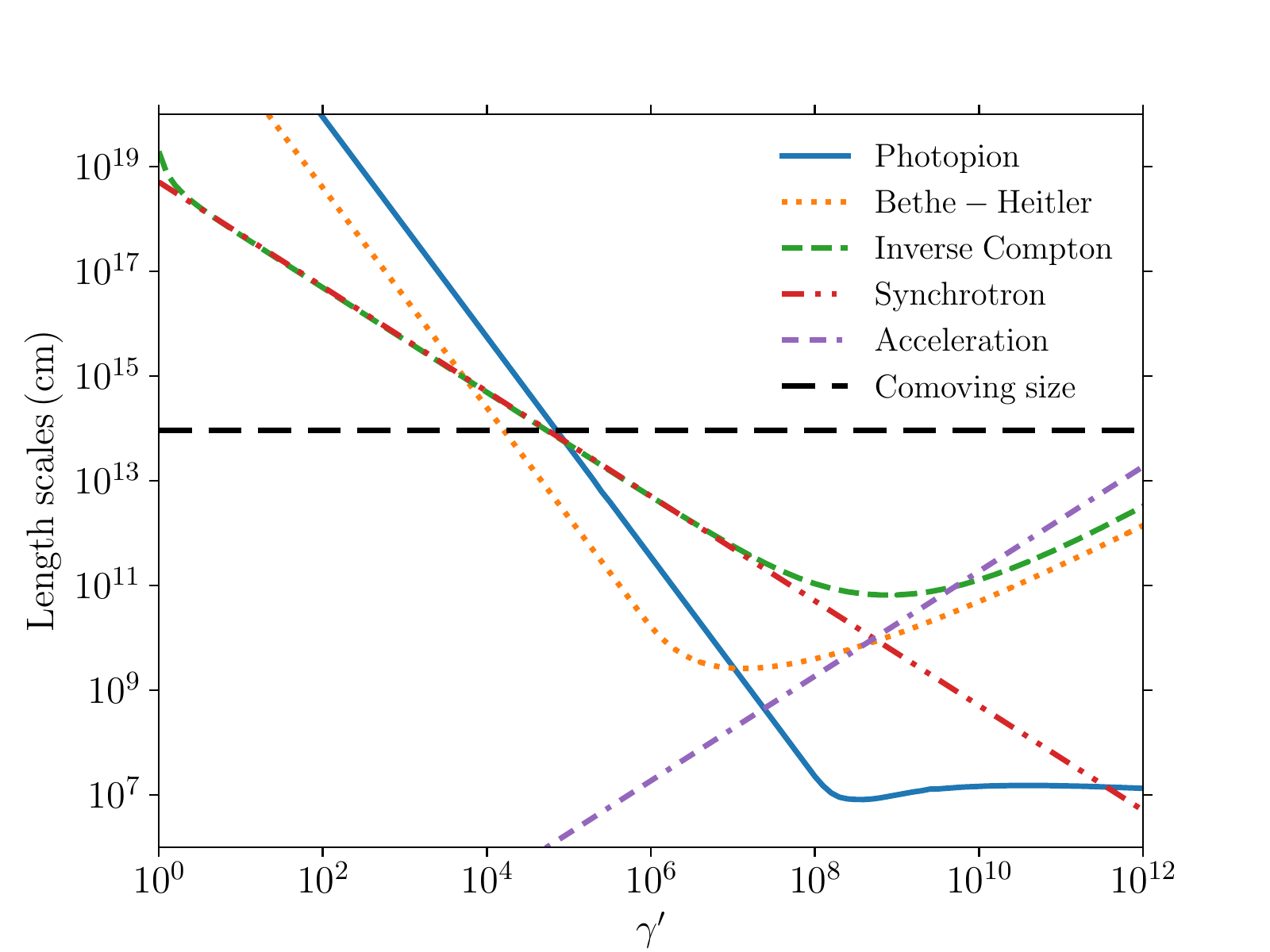}}
\end{minipage}
\begin{minipage}{\textwidth}
{\includegraphics[width=0.49\textwidth]{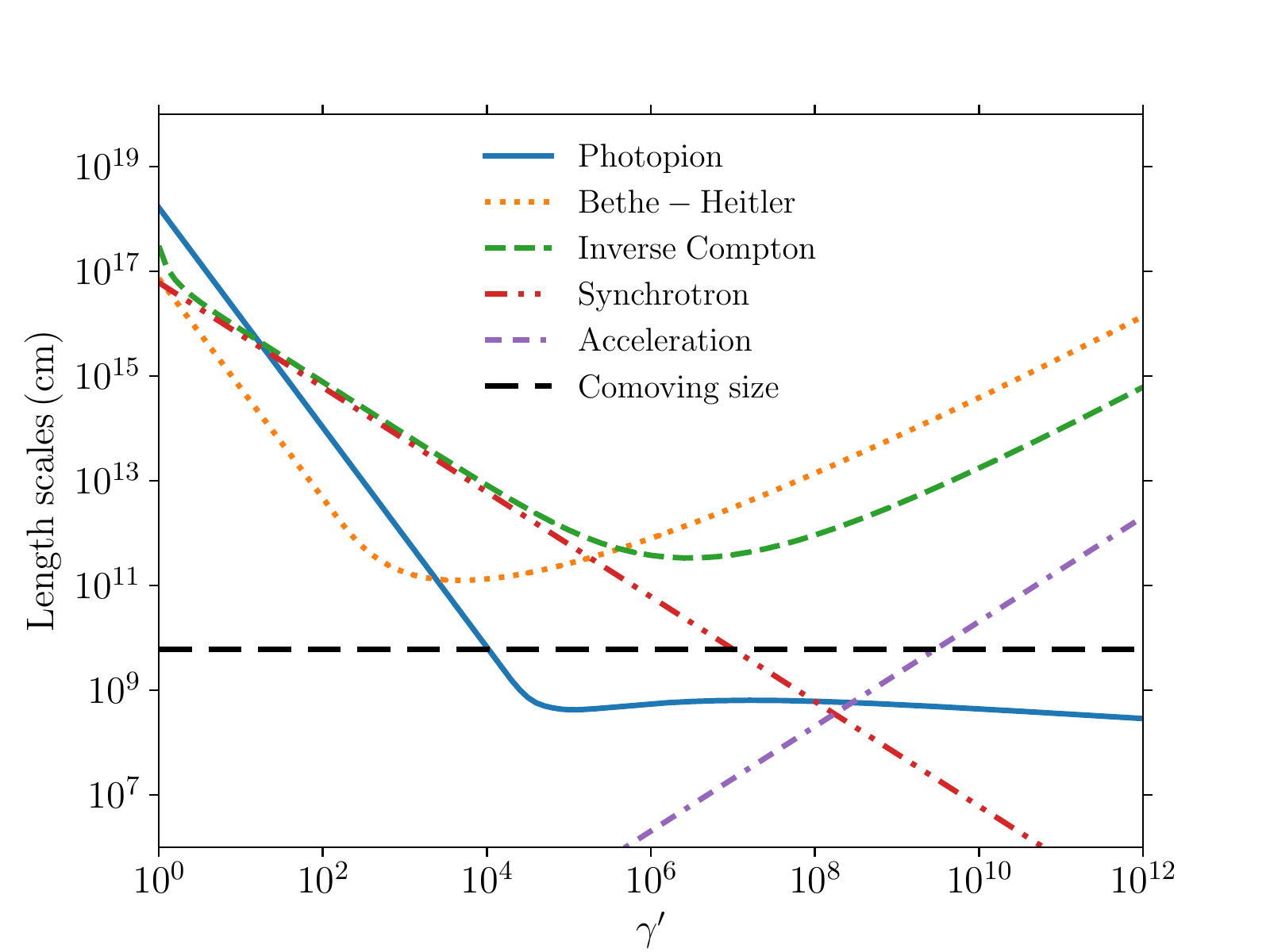}}
{\includegraphics[width=0.49\textwidth]{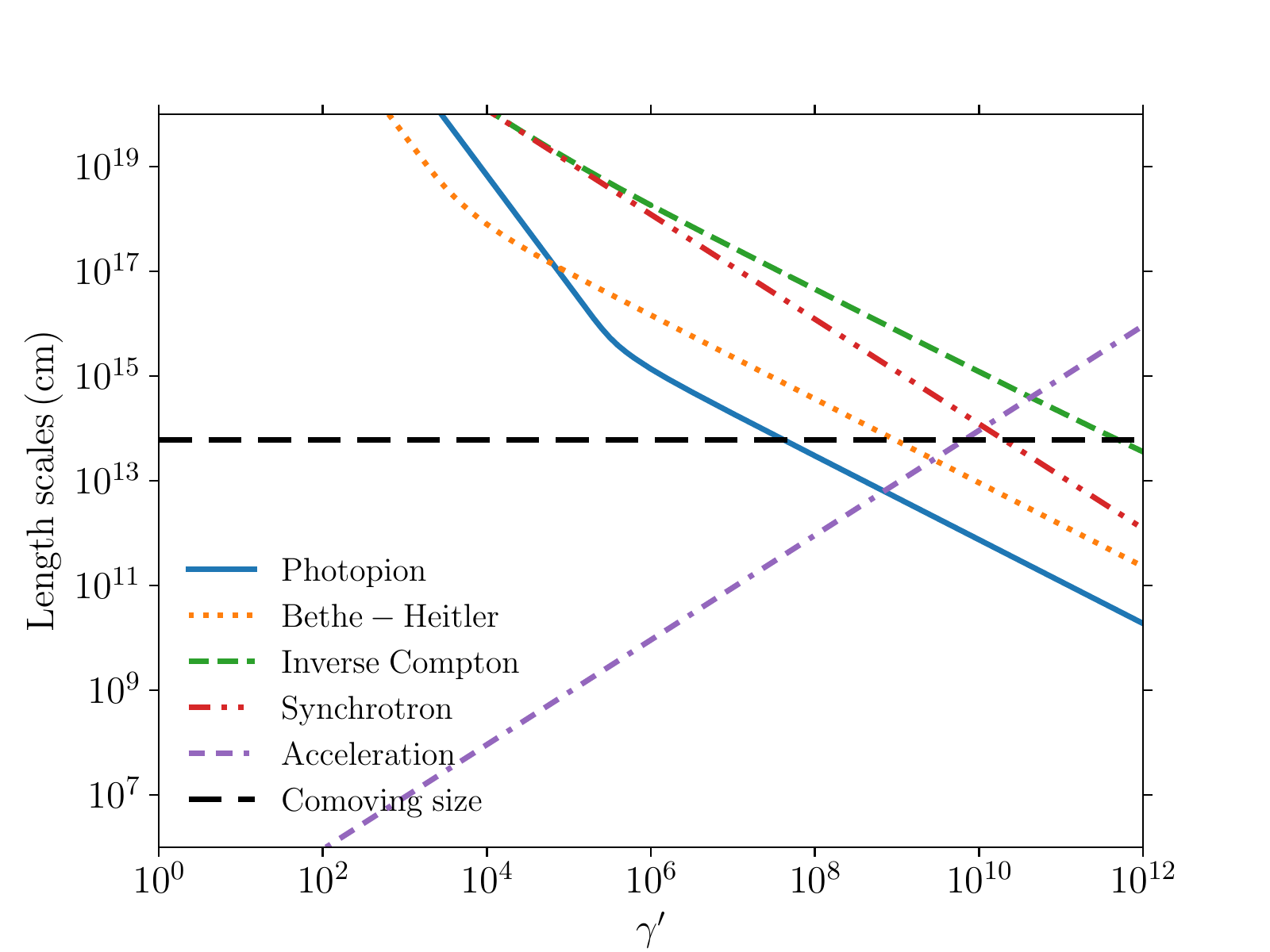}}
\captionof{figure}{Characteristic proton length scales as a function of the comoving Lorentz factor $\gamma'$, for the four case studies: millisecond magnetars with $B_{\rm p}=10^{14}\,{\rm G}$ (top left), and with $B_{\rm p}=10^{15}\,{\rm G}$ (top right), magnetar giant flares (bottom left) and tidal disruptions (bottom right). }\label{fig:MFP}
\end{minipage}

The shape of the photopion scattering length scale $l'_{p\gamma}$ is related to the photon spectrum. For low proton energies, the contribution of the high-energy part of the photon spectrum is dominant. For high proton energies, the low-energy part of the photon spectrum dominates. The slope $x_{p\gamma}$ of $l'_{p\gamma}(\gamma')$ depends on the slope of the photon spectrum, such as $x_{p\gamma}=1-x$ for a soft spectrum of index $x$ and $x_{p\gamma}=0$ for a hard spectrum.

The maximum energies of protons can be estimated by comparing acceleration and energy-loss length scales, by computing the minimum Lorentz factor such as $l'_{\rm acc}=l'_{\rm loss}$. The values shown in figure~\ref{fig:MFP} are consistent with the proton spectra from figure~\ref{fig:spec_cases}. Similar acceleration and energy loss length, together with decay length, can be computed for charged pions and muons.

\end{document}